 \documentclass{aastex}

\begin{document}

\title{SCALE-ADAPTIVE FILTERS FOR THE DETECTION/SEPARATION OF COMPACT SOURCES}

\author{D. Herranz$^{1,2}$, J.L. Sanz$^{1}$, R.B. Barreiro$^{3}$
and 
E. Mart\'\i nez-Gonz\'alez$^{1}$}

\affil{(1) Instituto de F\'\i sica de Cantabria,  
             Fac. de Ciencias, Av. de los Castros s/n, \\
             39005-Santander, Spain \\
 (2) Departamento de F\'\i sica Moderna, 
             Universidad de Cantabria,	     
             39005-Santander, Spain \\
 (3) Astrophysics Group, Cavendish Laboratory, 
Madingley Road, Cambridge
CB3 0HE, UK}

\begin{abstract}
This paper presents  scale-adaptive filters that optimize the 
detection/separation of compact sources
on a background. We assume that the sources have a multiquadric profile, i. e. 
$\tau (x) = {[1 + {(x/r_c)}^2]}^{-\lambda}, \lambda \geq \frac{1}{2}, 
x\equiv |\vec{x}|$, and a background modeled by a homogeneous and 
isotropic random field,
characterized by a power spectrum $P(q)\propto q^{-\gamma}, \gamma 
\geq 0, q\equiv |\vec{q}|$.
We make an n-dimensional treatment but consider two interesting 
astrophysical applications
related to clusters of galaxies (Sunyaev-Zel'dovich effect and X-ray emission).

\end{abstract}


\section{Introduction}

The detection of localized signals or features on one-dimensional (1D)
spectra or two-dimensional (2D)
images is one of the most challenging aspects of image analysis. 
Astronomical spectra
and images, for example, contain information that comes from a 
wide variety of physical
phenomena mixed with instrumental noises of 
diverse origin (white noise, $1/f$ noise,
etc.). Before the analysis of such images it
is necessary to carefully perform a {\it component
separation} in order to isolate the different physical 
sources that contribute to the 
data. In this work we are interested in optimizing the 
detection of localized sources with
spherical symmetry  and spatial profiles belonging to a given 
family of mathematical functions. In particular, 
we will consider the well-known
$\beta$ profiles of galaxy clusters
that are broadly used in X-ray and microwave Astronomy.
By `detection' we mean the determination of the position of the sources as well
as the estimation of parameters such as 
 the intensity at the central pixel ({\it amplitude}, hereafter) and
the characteristic scale of each source. 
Given the radial profile of the cluster,
knowing the amplitude and the scale is equivalent to knowing the total flux
at the observation frequency.
Therefore, the whole `detection' problem includes two different
challenges: on the one hand, how to maximize the probability of finding
a source against the noisy background in which it is embedded; on 
the other hand, how to accurately estimate the desired source 
parameters. Another related problem is how to distinguish between
`true' detections (i.e. those which correspond with real objects
in the data) and spurious ones. 

Given several images at different frequency channels,
it is possible to
use the knowledge about the frequency dependence of the different
components as well as 
their statistical properties to separate them. 
Wiener filtering (WF, Tegmark and Efstathiou 1996; Bouchet et al. 1999) 
and Maximum 
Entropy Method (MEM, Hobson et al. 1998, 1999) are 
powerful tools for component separation based on this idea. When the 
frequency dependence of the sources is not known 
the previous methods are inefficient. Moreover, if
only a single image is provided, the knowledge of the $\nu$-dependence 
of the different components is of no use. In that case, only  spatial 
properties (such as characteristic scale, profile, structure,
distribution and other statistical properties) might be used 
to perform the separation.
In the case we are going to consider in this work,
we intend to detect localized sources (that is, with a 
`small' characteristic scale)
with spherical symmetry and a given radial profile. 
All the other components in the image will be considered as 
a background (`noise') that should be removed in order to
optimize the detection of the signal.

The classical treatment to remove the background has been {\it filtering}.
A classical, maybe somewhat engineer-oriented definition of filter is:
`a filter is a device in the form of a piece of physical hardware
or computer software that is applied to a set of noisy data in order to
extract information about a prescribed quantity of
interest' (Haykin 1996). 
This reflects the intuitive idea of a filter: a tool that,
provided a certain data input, gives an output that have some
desirable properties. 
From the signal processing point of view, a filter is a kind
of \emph{system}, that is, a process that results in the transformation
of a signal. 
From the mathematical
point of view, a filter is an \emph{operator}:
\begin{equation}
L: f(\vec{x}) \longrightarrow g(\vec{x})=Lf(\vec{x})
\end{equation}
where $f$ is the input, $g$ is the output and $\vec{x}$ is the 
n-dimensional independent
variable.
Here we are interested in operators (filters) 
that are \emph{linear} and that have
\emph{translational invariance}, that is, that
 if the input is translated by $\vec{p}$ the output is also
translated by $\vec{p}$, that is, $g(\vec{x}-\vec{p})=Lf(\vec{x}-\vec{p})$.
A well-know result of signal processing theory is that
any linear operator (filter) with translational invariance is equivalent to
a convolution of the data with a certain function(\emph{convolution
property}). Any convolution can be expressed as a product
in Fourier space. Thus, linear operators (filters) with 
translational invariance  are frequency-selective (in the Fourier sense)
devices:
a filter selects or removes some of the features of the image
depending on their frequency or, if the independent variable has
dimensions of length, their \emph{scale}. 
For example, Gaussian filters have been thoroughly
used to suppress white noise from astronomical images.
Low and high-pass
filters are efficient in removing the small and the large 
features of an image, respectively, but are extremely
inefficient in dealing with localized sources since many
waves are required to represent such sources in Fourier space. 

Wavelet formalism is well suited to deal with localized
signals. The localized bases used in wavelet analysis allow one to obtain
a precise representation of local objects both in spatial
and frequency domains. In the context of
signal processing, wavelets can be considered
as a subset of the set of band-pass filters, capable of selecting 
a finite range of frequencies (scales) of an image, 
with the particularity that they are
characterized 
not only by a translation but also by a scaling. This scaling
allows one to perform a multiresolution analysis. 
The Mexican Hat wavelet (MHW) has been successfully applied to real and
simulated X-ray images (Grebenev et al. 1995, 
Damiani et al. 1997, Valtchanov et al. 2001, 
Freeman et al. 2001)
in order to detect X-ray sources, as well as to detect and extract
point sources from simulated microwave maps
(Cay\'on et al. 2000, Vielva et al. 2001a, 2001b).
Another kind of wavelet that has been used in X-Ray Astronomy is 
the difference of two Gaussians (Rosati et al. 1995, Vikhlinin et al. 1998,
Lazzati et al 1999).

We may wonder if a given wavelet such as the Mexican Hat
or the difference of two Gaussians is the best possible choice
in every case or if, on the contrary, there are different families of
functionals (operators), 
wavelets or not, which are better suited to each particular case.
It is clear that such an operator (filter) should take into account the 
shape (profile) of the source, its characteristic scale
and the statistical properties of the background in which is
embedded. 
In this context, the application of matched filtering (MF) was
recognized to be an effective method  for detection of faint sources (see
for example Irwin 1985). MF has been applied to the detection of 
faint sources in ROSAT PSPC images (Vikhlinin 1995).
Regarding CMB data analysis,
Tegmark \& Oliveira-Costa 
(1998) introduced a matched filter that minimizes the variance of the 
filtered map. Their filter takes maximum advantage of 
the knowledge of the source profile and the background properties, 
but it does not explicitly 
require that the detection has to be optimal at a fixed scale,
i.e., it is not optimal to detect sources of a given scale {\it and
only them}. On the other hand, Sanz et al. (2001) introduced 
a family of filters (pseudofilters or {\it scale-adaptive} filters) 
that produce maximum
detection {\it at a particular scale}. Optimal scale-adaptive
filters
have been successfully used to detect and extract point
sources from simulated microwave 
time ordered data sets (Herranz et al. 2002).

Sanz et al. (2001) showed that the MHW is nearly optimal 
to detect point sources convolved with a Gaussian beam
because of the relation between the MHW and the
laplacian of the Gaussian. If the source profile
is other than Gaussian, however, the optimal filter 
would be different. 
For example, an astronomical image may contain different
kind of objects with different spatial profiles. In that case,
it would be necessary to use different filters to detect
each kind of objects.
In this work we will apply 
optimal scale-adaptive filters to the separation and
detection of compact sources with a multiquadric
profile. Such profile is quite common in X-Ray and
CMB images.
 In section~\ref{adapt} we briefly describe the 
formalism of scale-adaptive filters and introduce 
the multiquadric profiles that we are going to study
as well as the concept of `gain'. Section~\ref{others} 
introduces other related filters such as the MHW and
the matched filter. 
A more extensive discussion about the differences between
scale-adaptive filters and wavelets is presented in this section.
In section~\ref{simulations} the
scale-adaptive filters are applied to simple simulations of
multiquadric profiles embedded in homogeneous and isotropic
backgrounds. 
Section~\ref{sec_compa} deals with an empirical comparation
between the scale-adaptive filters and two examples of wavelets,
namely the Mexican Hat Wavelet and the difference of two Gaussians.
Finally, in section~\ref{conclusions} we discuss our
conclusions.

\section{The scale-adaptive filter} \label{adapt}
\subsection{The concept of scale-adaptive filter}

Let us consider an n-dimensional (nD) image with data values defined by
\begin{equation}
y(\vec{x}) = s(x) + n(\vec{x}),\ \ \ x\equiv |\vec{x}|,
\end{equation}
\noindent where $\vec{x}$ is the spatial coordinate and $s(x)$
represents a compact source with spherical symmetry  and 
a characteristic scale (e.g. a single maximum at its
center and decay at large distances)
placed at the
origin. The background $n(\vec{x})$ is
modeled by a homogeneous 
and isotropic random field with average 
value $<n(\vec{x})> = 0$ and power spectrum $P(q), 
q\equiv |\vec{q}|$: $\langle n(\vec{q})n^*(\vec{q'})\rangle =
 P(q)¨\delta_D^n (\vec{q} - \vec{q'})$,
$n(\vec{q})$ is the Fourier transform of $n(\vec{x})$ 
and $\delta_D^n$ is the nD Dirac 
distribution. 

The idea of an optimal pseudo-filter (or a \emph{scale-adaptive 
filter}) 
has been 
recently introduced by the authors (Sanz et al. 2001).
Let us consider a generic spherically-symmetric filter,
$\Psi (\vec{x}; R, \vec{b})$, dependent on
$n + 1$ parameters ($R$ defines a scaling 
whereas $\vec{b}$ defines a translation)
\begin{equation} \label{genfilter}
\Psi (\vec{x}; R, \vec{b}) = \frac{1}{R^n}\psi \left( 
\frac{|\vec{x} - \vec{b}|}{R} \right).
\end{equation}
\noindent This generic filter is very similar to a 
continuous wavelet transform, i.e. it is characterized by 
self-similarity (or
covariance under $\vec{x} \rightarrow R(\vec{x}-\vec{b})$ 
transformation). In fact, $\Psi$ may be a wavelet if the wavelet
transform conditions (see section~\ref{wav_fil})  are met. Let us
consider not only wavelets but 
the whole set of filters that can be described by
 eq. (\ref{genfilter}).
We define the filtered field as
\begin{equation} \label{ffield}
w(R, \vec{b}) = \int d\vec{x}\,y(\vec{x})\Psi (\vec{x}; R, \vec{b}).
\end{equation}

Now, we are going to express the conditions in order to obtain an
 optimal (in the sense that will be described below) scale-adaptive 
filter for the
detection of the source $s(x)$ at the origin taking 
into account the fact that the source is
characterized by a single scale $R_o$. 
By `optimal' we mean that the filter must satisfy the following conditions:
(i) $ \langle w(R_o, \vec{0}) \rangle = s(0) \equiv A$, 
i. e. $w(R_o, \vec{0})$ is an \emph{unbiased} estimator 
of the amplitude of the source,
(ii) the variance of $w(R, \vec{b})$ has a minimum 
at the scale $R_o$, i. e. it is an 
\emph{efficient} estimator of the amplitude of the source, and
(iii)  $w(R, \vec{b})$ has a maximum at $(R_o, \vec{0})$.
The meaning of these conditions is simple: in order to maximize the
probability of detection one wants to reduce as far as possible
the contribution of noise (condition (ii)) while preserving the
sources (condition (i)). Additionally, condition (i) 
provides a normalization that gives directly the amplitude 
of the sources after filtering. Conditions (i) and (ii) alone
give birth to the so-called \emph{matched filter}. It 
is a well-known result of signal analysis 
that matched filter produces the maximum gain in SNR when
going from real to filter space.
By introducing further conditions in the making of the filter we constrain
the space of functionals from which the filters are drawn, thus
reducing the final SNR gain of the filter. Nevertheless, we
introduce a third condition in our definition of the `optimal filter'.
This last condition is set in order to optimize the
characterization of the scale of the source by the filter. 
As we will see
in further sections, condition (iii) allows one to 
establish a straightforward relation between the filter parameters and
the characteristic scale of the sources to be detected. Besides, 
it helps to reduce the probability of false detections.
A filter satisfying conditions (i) to (iii) is optimal for 
finding the sources, determining their scales and amplitudes and
discarding spurious detections \emph{at the same time}.

The filter satisfying  
conditions (i), (ii) and (iii) is given by the equation
\begin{equation} \label{SAFmq}
\tilde{\psi} (q) \equiv \psi (R_oq) = \frac{1}{\alpha \Delta } 
\frac{\tau (q)}{P(q)}
\left[nb + c - (na + b)\frac{dln\tau}{dlnq}\right],
\end{equation}
\begin{equation}
\Delta \equiv ac - b^2,\ \ \ \alpha \equiv 
\frac{2{\pi}^{n/2}}{\Gamma \left(\frac{n}{2}\right)}, 
\end{equation}
\begin{equation} 
a\equiv \int dq\,q^{n - 1}\frac{{\tau}^2}{P},\ \ \ 
b\equiv \int dq\,q^n \frac{\tau}{P}\frac{d\tau}{dq},\ \ \
c\equiv \int dq\,q^{n + 1}\frac{1}{P}{\left(\frac{d\tau}{dq}\right)}^2.
\end{equation}

\noindent where $\tau$ is the profile of the source in Fourier space ($s(q) = A\tau (q)$).
Generically, $\Psi$ is not positive.
Moreover, as mentioned above 
$\Psi$ does not define in general a continuous wavelet transform,
although it has its self-similarity property. We could obtain a 
continuous wavelet transform in the same way described above simply
by introducing in the derivation of the filter 
the additional conditions that a wavelet transform must satisfy.
However, this would lower the SNR gain.
This would
reduce the usefulness of the filter from the point of view
of pure detection. Therefore, we will not introduce any other
condition. 

The filter given by eq. (\ref{SAFmq}) \emph{adapts}
to the source profile, the background and the scale of the source, 
i. e. the name ${\it
scale-adaptive}$ filter.
Taking into account equation (\ref{ffield}), the amplitude is estimated as
\begin{equation} \label{eq_amp}
A = w(R_o, \vec{0}) = \int d\vec{q}\,y(\vec{q})\tilde{\psi} (q),
\end{equation}
\noindent whereas an estimation of the error is given 
by the dispersion $\sigma_w$ of the
filtered image
\begin{equation} \label{eq_sigma}
\sigma_w = \left[\langle (w(R_o, \vec{b}))^2\rangle - A^2\right]^{1/2},
\end{equation}
\noindent with the average including all points $\vec{b}$ in the image.
This equation provides a theoretical estimation of the variance 
of the estimation of the amplitude of sources.

The previous equations have been used by the authors to obtain 
the scale-adaptive filter for a Gaussian
and an exponential profile (Sanz et al. 2001).

\subsection{Multiquadric profile}
In some astrophysical/cosmological applications the source is modeled 
by a multiquadric, i. e. the profile is given by
\begin{equation} \label{multiquadric}
\tau (x) = \frac{1}{ \left[ 1 + 
\left(\frac{x}{r_c}\right)^2 \right]^{\lambda}},\ \ \ \lambda \geq 1/2.
\end{equation}

\noindent Typical examples are, for the 2D case, 
the emissions in the microwave and X-ray bands
with $\lambda = \frac{3\beta - 1}{2}, \frac{6\beta - 1}{2}$, 
respectively, for a $\beta$-profile
for the electron number density 
$n_e(r)\propto {[1 + {(x/r_c)}^2]}^{-\frac{3}{2}\beta}$. 
Assuming the standard value 
$\beta = 2/3$ one trivially obtains $\lambda = 1/2, 3/2$ for
the microwave and X-ray emissions, respectively.

Assuming a scale-free power spectrum $P(q) = Dq^{- \gamma}$, the equations 
(5-7) lead to
the filter
\begin{equation}
\tilde{\psi} (q) = \frac{1}{\alpha a^{\prime}}\frac{\Gamma (\lambda )}
{2^{1 - \lambda }(\gamma + n)}{(qr_c)}^{\gamma + \lambda - \frac{n}{2}}
\left[P\,K_{\lambda - \frac{n}{2}}(qr_c) + 
Q\,qr_cK_{1 + \lambda - \frac{n}{2}}(qr_c)\right] ,\ \ \ 
\end{equation}
\begin{equation}
P\equiv 2\gamma - (n - \gamma )(\gamma + 2\lambda ),\ \ \  
Q\equiv 2(n - \gamma )\frac{\gamma + 2\lambda + 1}{\gamma + 4\lambda - n},
\end{equation}
\begin{equation} \label{aprime}
a^{\prime}\equiv \frac{2^{\gamma + 2\lambda -3}}{\Gamma \left(\gamma 
+ 2\lambda \right)}\Gamma \left(\frac{\gamma + n}{2}\right)
\Gamma^{2} \left(\frac{\gamma 
+ 2\lambda}{2}\right)\Gamma \left(\frac{\gamma + 4\lambda - n}{2}\right).
\end{equation}
\noindent In the previous equations $\Gamma$ denotes the Gamma function whereas $K$
denotes the Bessel $K$ function. 
Table~\ref{filtertab1} gives the analytical form of 
the scale-adaptive filter for $\lambda = 1/2, 3/2$ and
different values of the spectral index $\gamma = 0, 1, 2, 3$ on 
Fourier space and real space.
The 2D scale-adaptive filters for the same parameters are shown
in the left side of figure~\ref{filters}.

\subsection{Detection level and gain}

One can define the detection level on real ($D$) and filtered
($D_w$) space as
\begin{equation}  \label{gaineq}
{\mathcal{D}} = \frac{A}{\sigma},\ \ \ {\mathcal{D}}_w = \frac{A}{{\sigma}_w},
\ \ \  g = \frac{\sigma}{{\sigma}_w},
\end{equation}
\noindent where $\sigma , \sigma_w$ are the dispersion in the real
image and filtered image, respectively.
The gain $g$ gives the amplification of the sources achieved in
filter space with respect to the original image.

The most straightforward way to identify sources in an image is to
look for  peaks of the signal above a certain threshold. In Astronomy is
usual to use thresholds that are a certain number of 
times the dispersion of
the image.  Therefore, 
the number of faint sources that can be detected with a filter
is proportional to the gain of the filter.
As mentioned above,
conditions (i) and (ii) alone (corresponding to a matched filter) 
guarantee the maximum possible
gain; if additional conditions are required the gain will
be lower. 

Other issue that has to be taken into account
is the \emph{reliability} of the detection. 
The identification of sources as peaks above a certain 
threshold (e. g. $3\sigma_w$) in real or in 
filter space gives a finite probability of false detections
due to noise fluctuations. If the noise is Gaussian and
uncorrelated it is easy to control the false discovery rate
simply by choosing the corresponding $\sigma$ level threshold. Unfortunately,
this is not true in many cases. In general, the relation between
the value of the threshold and the false discovery rate is not
straightforward and must be calibrated in some way, for example using
simulations. A compromise between sensitivity and reliability must
be reached. A way to overcome this problem is to use additional information
about the sources in order to distinguish them from the noise 
fluctuations. In many cases, one looks for objects that have a
fixed scale in the image. As an example, point sources in CMB maps
appear as Gaussian-like features (due to the beam psf) with
the scale of the beam width. Scale-adaptive filters are
specifically designed to take advantage of the characteristic scale
of the sources in order to reduce spurious detections. By means of
condition (iii), only features that have the same 
 characteristic scale of 
variation as the sources we intend to detect are enhanced. 
In other words, the gain experimented by 
a background fluctuation drops quickly when it
has a characteristic scale of 
variation different from   
that of the sources.

\section{Other filters} \label{others}

For comparison with the filters developed in the previous section, 
we shall briefly introduce
other filters that have been extensively used 
in the literature. Two of them, the Mexican Hat
and the difference of two Gaussians, are wavelets.
The other is the well-known  `matched' filter (MF).

\subsection{Wavelets and filters} \label{wav_fil}

The development of wavelet techniques applied to signal processing has been 
very fast in the last ten years.
A wavelet is a `small wave' which has its energy concentrated
in space and allows simultaneous space and scale (frequency) analysis based on
location and similarity.
Wavelets are so useful because of their good space-frequency localization.
Besides, almost all useful wavelet systems allow one to reconstruct a
given signal in terms of a wavelet basis that  is generated
from a single scaling function by simple \emph{scaling} and
 \emph{translation}. Eq. (\ref{genfilter}) shows these two basic operations. 
There is not a unique choice for the wavelet basis.
The ability of giving better approximations to a signal by means of
successive `refinements' of the wavelet is called the 
\emph{multiresolution} condition. In order to have a wavelet
(i.e. to allow a multiresolution
analysis and the reconstruction of signals), the functions in eq. 
(\ref{genfilter}) must satisfy a number of additional conditions,
namely the condition $\int d\vec{x} \ \Psi=0$ and
the `admissibility'
condition $\int dq \ q^{-1} \psi^2(q) < \infty$. 
Due to their good space-frequency localization and their ability to
perform a complete decomposition of the signal,
wavelets are ideal tools for identifying features in images, denoising
and data compression.

From de point of view of mere source detection, however, not all 
the properties of wavelets are equally necessary. Clearly, good
space-frequency localization is an important condition to be met. On the
contrary, a full decomposition and posterior reconstruction of the image
is not necessary, since we are only interested in 
finding the position of
the sources and a handful of parameters such as the amplitude; 
as explained in the previous section, each new condition to be satisfied
by a filter reduces the SNR gain in the output.
In this context, the filter (\ref{SAFmq})
resembles a wavelet in the sense of space-frequency localization,
but is more generic and therefore it will produce higher
gains (and, therefore, will be more sensitive to weak sources).

A different approach to the detection problem is the denoising
of images previous to the actual detection. Since noise usually
manifests at scales different from the source characteristic scale,
scale-adaptive filters are good at denoising. Wavelet-based
denoising algorithms are also very powerful and efficient
(as an example applied to CMB data analysis see Sanz et al. 1999a,b).
Wavelet-oriented denoising techniques 
are out of the scope of
this work (for an introduction to this topic see for example
Odgen 1997).

\subsubsection{The Mexican Hat wavelet}

The well-known Mexican Hat wavelet is defined by
\begin{equation} \label{mexhateq}
\Psi (\vec{x}; R, \vec{b}) = \frac{1}{R^n}\psi \left(\frac{\left|
\vec{x} - \vec{b}\right|}{R}\right),\ \ \ 
\psi (x) \propto (n - x^2) e^{- x^2/2}              
\end{equation} 
\begin{equation}
\psi (q) \propto q^2e^{- \frac{1}{2}q^2}.
\end{equation}
\noindent This type of wavelet has been extensively 
used for point source detection. Optical
images of galaxy fields have been analyzed to detect voids and high-density structures in the
first CfA redshift survey slice (Slezak et al. 1993). Microwave 
images have been analyzed (Cay\'on et al. 2000; Vielva et al. 2001a) 
and combined with the maximum
entropy method (Vielva et al. 2001b) to obtain 
catalogs of point sources from simulated maps
at different frequencies that will be 
observed by the future Planck mission. On the other hand, 
the MHW has also been used to detect X-ray sources (Grebenev et al. 1995,
Damiani et al. 1997) 
and presently for
the on-going XMM-Newton mission (Valtchanov et al. 2001) and 
Chandra (Freeman et al. 2001).

The Mexican Hat Wavelet is well suited to deal with Gaussian
structures due to its relation with the Laplacian 
of the Gaussian. Intuitively, it gives good results because
its profile is highly correlated with the Gaussian profile. 
In fact, Sanz et al. 2001 showed that the scale-adaptive filter
for a two-dimensional Gaussian source embedded in a noise
characterized by the power spectrum $P(q) \propto q^{-\gamma}$
with $\gamma=0$ (white noise) and $\gamma=2$ coincides with
a Mexican Hat Wavelet. If the source profile is not
Gaussian or the background is not described by the relation above,
the MHW is not optimal.

\subsubsection{Difference of two Gaussians}

A wavelet kernel can be constructed by subtracting two Gaussians.
For example, in the two-dimensional case
\begin{equation} \label{2gauss}
\Psi (\vec{x}; R, \vec{b}) =
\frac{1}{R^2}\psi \left(\frac{\left|
\vec{x} - \vec{b}\right|}{R}\right),\ \ \ 
\psi(x) =
2\left[e^{-x^2}-\frac{1}{2}e^{x^2/2}\right].
\end{equation}
This wavelet shows a positive core and a outer negative ring,
and the convolution with any linear function $s(x,y)=ax+by+c$ is
zero (this is also true for the Mexican Hat). Therefore, any slowly varying
background that can be locally approximated by such kind of
linear functions is subtracted by this wavelet. The difference of
two Gaussians has been applied to detect X-ray sources 
(Rosati et al. 1995, Vikhlinin et al. 1998,
Lazzati et al 1999).

\subsection{The matched filter}

If one removes condition (iii) defining the ${\it scale-adaptive}$ filter in
the previous section, 
another type of filter can be found after minimization of the variance
(condition(ii)) with the constraint (i)
\begin{equation} \label{matched}
\tilde{\psi}_m (q) = \frac{1}{\alpha a}\frac{\tau (q)}{P(q)}.
\end{equation} 
\noindent 
This is usually called a ${\it matched}$ filter.
In general, the ${\it matched}$ and ${\it scale-adaptive}$ 
filters are different. In the
former case, one obtains a slightly larger gain (although 
for $\gamma \geq 0.5$ the gain is the same from the practical point of
view, i.e. the relative difference is less than $20\%$) 
but sources must be identified `a posteriori' with an extra criterion whereas
the ${\it scale-adaptive}$ filter allows one to get the sources in a
straightforward manner. Thus, from the
methodological point of view and reliability (detection of spurious sources) 
it is better to use
the ${\it scale-adaptive}$ filter unless the image is 
completely dominated by white noise.

   Matched filters have been used recently to detect clusters 
of galaxies from optical 
imaging data (Postman et al. 1996; Kawasaki et al. 1998)
and sources from X-ray images (Vikhlinin 1995). In this 
approach the method uses
galaxy positions, magnitudes and photometric/spectroscopic 
redshifts if available to find
clusters and determine their redshift.

For the case of a source profile given by eq. (\ref{multiquadric}) 
and a scale-free power 
spectrum given by $P(q)\propto q^{-\gamma}$, the previous formula 
(\ref{matched}) 
leads to the 
following matched filter
\begin{equation} 
\tilde{\psi}_m (q) = 
\frac{1}{\alpha a^{\prime}}\frac{\Gamma (\lambda )}{2^{1 - \lambda }}
{(qr_c)}^{\gamma + \lambda - 
\frac{n}{2}}K_{\lambda - \frac{n}{2}}(qr_c),     \ \ \ 
\end{equation}

\noindent with $a^{\prime}$ given by eq. (\ref{aprime}). 
For $\lambda = 1/2$ the matched filter is 
not defined for white noise (i.e. $\gamma = 0$), 
whereas for other values of $\gamma$ is given by 
table~\ref{filtertab2} in Fourier space and real space,
respectively. The same filters appear in the right 
side of figure~\ref{filters}.
Note that the scale-adaptive filters reach their peak 
around $qr_c \sim 1$, whereas there is more 
dispersion in the position of the peak of the matched filters. 
This means that scale-adaptive filters
enhances the signal against the background in
  q-space at scales comparable with the characteristic scale
  of the source.
We remark that for $\gamma = n$ the scale-adaptive filter and the 
matched filter coincide.
For the $\lambda=1/2$ and 
$\gamma = 0$ case, we have assumed the modified profile
\begin{equation} 
\tau (x) = N[\frac{1}{{(r^2_c + x^2)}^{1/2}} - 
\frac{1}{{(r^2_v + x^2)}^{1/2}}], \ \ \
N\equiv \frac{r_cr_v}{r_v - r_c},
\end{equation}

\noindent where $r_v$ is a cut-off radius. 
The behavior of this profile is: $\tau (0) = 1$ and
$\tau (x)\propto x^{-3}$ for $x\gg r_v$.

\section{Numerical simulations: application and results} \label{simulations}

In order to show the performance of the scale-adaptive filters we 
have simulated realizations of multiquadric profiles embedded
in backgrounds of the type $P(q)=Dq^{-\gamma}$. For the sake of simplicity
we simulate ideal multiquadric profiles, not taking 
into account the effect of the detector beam.
However, the method can be generalized to include this effect
by modifying the input source profile.
We also consider only integer values of $\gamma$, so we can 
directly use the filters from
Table 1. In a more realistic case, where the power spectrum does not follow
such a simple law, it could be directly estimated from the data and
the filter numerically calculated.

\subsection{Microwave emission and the SZ-effect of clusters}

\subsubsection{SZ clusters of equal size} \label{sec_sim1}

One of the most promising applications of scale-adaptive filters is the 
detection and extraction of the emission of galaxy clusters 
due to the Sunyaev-Zel'dovich (SZ) effect in microwave 
maps. Maps of the Cosmic Microwave Background (CMB) contain contributions
from a variety of foregrounds (the SZ effect among them)  
and different types of noise. If we approximate the power spectrum 
of a typical CMB map by a $P(q)=Dq^{-\gamma}$ law,
the effective index $\gamma$ ranges between values near $0$ (for regions
dominated by white noise) to $\sim 3$ (for regions dominated by
dust emission). As an example, we first simulated a $512\times512$ pixel field
containing 100 randomly distributed 
`clusters' with a multiquadric profile ($\lambda=1/2$).
All the `clusters' have the same scale
($r_{c}=1.0$ pixel) and amplitudes distributed between 0.1 and 1 (in arbitrary units).
With a convenient rescaling of the amplitude, this could
simulate, for example, a $12^{\circ}.8 \times 12^{\circ}.8$ field of the sky filled with 
clusters of several arcmin of extent.
The simulated clusters are shown in the left panel of figure~\ref{sim1}. 
A $P(q)=Dq^{-3}$ background was added so that the peaks of sources are 
{\it on average} at the $2\sigma$ level of the final map. The map containing
the sources and the background (`noise') is shown in the center of
figure~\ref{sim1}. In the following, we will call this simulation 
`simulation 1'.

Following table 1, the scale-adaptive filter for the case $\lambda=1/2$, 
$\gamma=3$ is
$\tilde\Psi_{o}(q)=4(3-y)y^{2}e^{-y}/3\pi$, $y=qr_{c}$. 
The result of applying the scale-adaptive filter on the simulated field is
shown in the right panel of figure~\ref{sim1}.
For the sake of clarity, only the pixels above the 
$3\sigma$ level have been plotted. 
After filtering, the detection is performed by looking
for peaks above a certain threshold. 
Knowing the initial position and amplitude of the simulated sources,
it is possible to determine quantities such as the mean error in the
estimation of the source parameters as well as the reliability of the 
filter (that is, the probability of detecting spurious `sources').
The results are summarized in 
table~\ref{tb3}. 
The first column in table~\ref{tb3} shows the $\sigma$ 
threshold used for the detection. 
The second column shows the number of true detections found. The third column
indicates the number of spurious  `sources' detected.
 The fact that above $4\sigma$ we
find more spurious detections than above the $3\sigma$ threshold may
seem surprising. This can be easily explained 
taking into account that our detection method looks for 
sets of connected pixels over
a given threshold. Therefore, a `lump' with two peaks may be seen as a single
source if the detection threshold is low enough whereas 
a higher detection threshold may 
 split the `lump' into separated peaks. 
The fourth column in table~\ref{tb3}
indicates the mean error in the determination of the position of the source in 
pixel units. As can be seen, all the detected sources were correctly located.
The mean relative bias given in the fifth column 
of table~\ref{tb3}
is defined as $\bar{b}_A (\%) = 100 \times \frac{1}{N}
\sum (A_{i}-A_{o_{i}})/A_{o_{i}}$, where $A_{o_{i}}$ 
and $A_{i}$ are the original and estimated amplitudes of the sources
respectively and $N$ is the number of considered sources.
The mean relative error given in the sixth column is defined as 
$\bar{e}_A = 100 \times \frac{1}{N}
\sum \mid A_{i}-A_{o_{i}} \mid /A_{o_{i}}$. The seventh column gives
the mean gain, as defined in eq. (13). 
The most striking fact in this case
is the high gain ($\sim 4.4$) that is found in all the cases. 
This is due to the fact that the scale-adaptive filter (which acts as a
band-pass filter) is very efficient in removing the large scale
structure that characterizes the $q^{-3}$ background.
The individual gain for each detection as a function of the
true amplitude of the source is shown 
in the bottom panel of figure~\ref{results1}. 
We see that there is a certain dispersion around the average value
(represented in figure~\ref{results1} as an horizontal dotted line) but,
in general, the gain is nearly independent of the original amplitude
of the source. Only for the
brightest and the faintest sources this independence is lost. Bright 
sources tend to have lower gains than the average and vice versa. 
This is related to what is seen in the 
top panel of figure~\ref{results1}, which
shows the estimated amplitudes of the detected sources versus the
original amplitudes. 
Between the original and estimated amplitudes there is a strong 
linear regression
but with a small positive deviation from the $y=x$ law. This
corresponds to the $\sim7\%$ 
relative bias that is given in table~\ref{tb3}. 
Again, both bright and faint 
sources deviate from this law.

Figure~\ref{results1} can be
explained considering the detection strategy we use. We look for peaks above a
certain threshold, that is, for local maxima in the filtered image.  
In an ideal case, the value of the peak gives directly the amplitude
$A$ of the sources (due to condition (i) of scale-adaptive filters).
Unfortunately, a certain amount of statistical residual noise is always found
in the filtered images. This noise is superimposed to the signal peaks
and produces an error in the estimation of $A$. This error can be positive
or negative. Since our selection criterion involves looking for local
maxima, positive errors are more likely to be detected than negative
ones. This leads to a systematic statistical positive bias.
This effect has more relevance for faint sources due to the fact 
that the relative contribution of the residual noise to the signal
is higher in this case.

There is one more reason for the higher gain in the case of faint sources.
The number counts of faint sources (near the detection level) is
affected by the well-known `detection bias'. Only the sources
that are more amplified than the average can be detected; therefore, 
the mean gain of weak sources is higher than the mean gain of bright 
ones.  From eqs. (\ref{eq_amp}) and (\ref{eq_sigma}) is
straightforward to show that the gain of the scale-adaptive filter does
not depend on the amplitude of the source. Therefore, the differences
among the gains obtained in different sources is caused by 
the fluctuations due to residual noise. 

As $A_{o}$ increases in the top panel of figure~\ref{results1},
the bias decreases and eventually drops to negative values. This suggests
the existence of a systematic effect that gives a negative bias and that
is compensated by the positive bias produced by the detection method when the 
sources are weak enough. Such
negative bias has been reported in relation with scale-adaptive filters in 
Sanz et al 2001 and Herranz et al. 2002, 
and is due to two kinds of effects. On the one hand, the
normalization given by condition (i) 
of scale-adaptive filters is calculated for
an ideal profile on an infinite, continuous field (the limits of the integral 
are $0$ and $\infty$), while the real images are finite and discrete
(pixelized).  
On the other hand, the correlation between the (analytic) filter
(\ref{SAFmq}) and the discrete, pixelized, `real' data is not 
perfect.

In spite of these small systematics, 
the performance of the scale-adaptive filter is very good in
the sense that with a simple application of a filter and a thresholding 
detection scheme we are able to recover a very significant number of 
sources and estimate their amplitudes with errors not larger than
 $\sim 10 \%$. The reliability of the method is such that over a $3\sigma$
threshold there are a $15\%$ of spurious detections.
A possibility to remove this small bias is to calibrate it
as a function of the amplitude of the sources by means of 
additional simulations using the same background and artificial
sources with known amplitudes.

\subsubsection{SZ clusters with different sizes} \label{sec_sim2}

In a more realistic case the scale $r_{c}$ of the sources will
not be known {\it a priori}. For example, in a CMB map, 
clusters of different scales will be present, going from almost point-like 
sources to large
structures that extend across several pixels. Under these conditions, 
it does not seem clear which one should be 
the scale of the scale-adaptive filter. 
Fortunately, condition (iii) of the optimal scale-adaptive filters implies
that the coefficient 
at the position of the source and at the `right' scale is a maximum. 
Therefore, the strategy to follow is to filter the image with a set of $N$
filters with different 
`tentative' scales $r_{c_{i}}$, $i=1,...,N$. For a given detection, 
the maximum value among the  
coefficients at the position of the source of the different filtered maps
 will correspond to the $r_{c_{i}}$ closer to the $r_{c}$ of the cluster.
This method is similar to 
a multiresolution analysis. The discrete wavelet transform (DWT) allows
one
to perform a multiresolution analysis by varying the scale of
the analyzing wavelet in logarithmic samples ($2^j$, $j=1,2,...,n$).
Such approach has been successfully used in X-ray data analysis (see 
for example Vikhlinin et al. 1998 and references therein). 
This allows an optimal decomposition of the signal in 
frequency space.
Our approach is more similar to a continuous wavelet transform (CWT),
in which the scale varies in a continuous way. Since we can not
calculate an infinite number of continuous scale variations, we
sample the scale space as densely as needed until we reach the
desired precision. 
Thus, the expected error in the determination of $r_c$ 
should be the size of the `step' in the sampling scales. In fact, 
the estimation of this error is more complicated and depends 
on the relative sizes of the core radius and the pixel, the 
characteristic scale of the background fluctuations and the shape
of the sources (for example the value of $\lambda$ will be relevant
for this matter). We suggest to estimate the expected errors by
means of simulations. As a first approximation, however, we
will consider through this work that the expected error is
simply the size of the `step' in the sampling scales.
 
To test this point we performed a new simulation (simulation 2, hereafter) 
with 100 `clusters' with $r_{c}$ distributed between 0.5 and 2.0 pixels and
amplitudes between 0.1 and 1.0 (in arbitrary units). 
The noise is similar to that of
simulation 1. The simulated map was filtered with scale-adaptive filters with
the parameter $r_{c}=$ 0.25, 0.50, 
0.75, 1.00, 1.25, 1.50, 1.75, 2.00, 2.25 and
2.50 pixels. 
In figure~\ref{sim2} we plot the original clusters (left panel),
the simulated map with noise (central panel) and the map 
filtered with the scale-adaptive
filter $r_{c}=1.50$ (right panel). The sources are detected over each
filtered map by selecting peaks above a certain threshold. If one of
such peaks is detected 
at the same position on different filtered maps, it is unlikely that 
it corresponds to pixel-scale noise (although it could be a fluctuation of 
the noise at a scale of the order of the source scale). 
Looking for detections in several 
filtered maps will help to reduce the number of
spurious sources in the output. 
For every detection that is 
present in several filtered maps 
we look for the 
maximum of the coefficients at the central position. 
In particular, we only consider sources that appear at least at 5 scales. This 
number was chosen because with 5 consecutive scales it 
is possible to `cover' the 
pixel size 
in the scale space and because it gives a good compromise between the number of detections and
the number of spurious detections, as will be seen through this section.
The maximum gives both the
scale of the source and its amplitude. 

The results are shown in table~\ref{tb4}.  
The main difference between tables~\ref{tb4} and~\ref{tb3} is the
lower gains
that are obtained in simulation 2. This result is expected
because the set of filters used only fit clusters with some
specific values of $r_{c}$. For
the rest of clusters the filters are only approximately optimal. Moreover, any
error in the determination of $r_{c}$ will lead to a wrong determination of
the amplitude. The top panel of figure~\ref{results2} 
shows the estimated $r_{c}$ versus the original
$r_{c_{o}}$ 
in simulation 2. There is a significant dispersion around the 
$r_{c}=r_{c_{o}}$ line (represented by 
a dotted line in the top panel of figure~\ref{results2}).
As shown in table~\ref{tb4}, the mean error in the determination of $r_{c}$ is 
$\sim 0.15$ pixels, with a similar bias towards 
estimating higher values of $r_{c}$ than the real ones. If we increase 
the `resolution' in $r_{c}$ by increasing the number of 
filters the error is not significantly reduced. This indicates that it is not
possible to determine scale parameters with a much better resolution than the
pixel scale. In spite of this limitation, the errors in the determination 
of the amplitude remain reasonable ($\sim 10 \%$).
This indicates that the estimated amplitude does not vary
significantly with $r_c$ in the neighborhood of the real value of the
core radius. Therefore, errors under the pixel scale in the
estimation of $r_c$ have little effect in the determination of the
amplitude. Lower panel of
figure~\ref{results2} shows the estimation of the amplitudes in simulation
2. 
As in simulation 1, a small positive bias is found.
The gain shows the same behavior with respect to the amplitude of
the sources as in simulation 1 
(an asymptotic decrease of the gain with the amplitude of the 
clusters) and is almost 
independent of the size of the clusters.
Small clusters tend to have slightly higher gains than large clusters. This 
could be due to the fact that small clusters are detected in maps filtered
with small $r_{c}$ parameter and therefore in these maps
the pixel-scale residuals are worse removed than in maps filtered with a 
large $r_{c}$.
Maps filtered with small $r_{c}$ show more contribution from 
pixel-scale noise residuals and therefore 
the peak amplitude is more likely to be overestimated.
The number of spurious detections drops to 0 due to 
the fact that we discard those `candidates' that are not detected in
at least 5 filtered maps.
Indeed a similar constraint is imposed in the MHW method used by
Vielva et al. (2001a and 2001b), who performed a 
 fit at several scales in order to estimate
the amplitude of point sources as well as to discard spurious
detections. Unfortunately, this constraint also reduces the number of true 
detections. This is an example of a well-known and unavoidable problem
in detection theory: the cost of reducing spurious detections is to reduce 
the number of true detections and, conversely, relaxing the 
selection criteria to include 
more true sources leads to an increase of spurious detections.

\subsection{X-ray emission and clusters}

Other straightforward application of scale-adaptive filters is in the field of X-ray
astronomy. X-ray emission from galaxy clusters roughly follows a multiquadric 
profile with $\lambda=3/2$. Unfortunately, real X-ray images suffer from 
strongly non-isotropic 
point spread functions that distort cluster profiles to quite odd and
asymmetric shapes. For this work, however, we will consider (as we did in
the previous case of SZ emission) that instrumental effects have
been somehow 
corrected and that we only have ideal clusters and noise. For simulation 3 we
distributed 100 ideal multiquadric profiles with $\lambda=3/2$,
$r_{c}$ between 2.0 and 4.5 pixels
and amplitudes between 0.1 and 1.0 (in arbitrary units). 
The dominating background in X-ray images is Poissonian shot-noise that,
when the exposure time is long enough, can be approximated by white noise.
We added white noise ($\gamma=0$) to our simulation
 so that, 
given the amplitudes of the
clusters (in their   arbitrary units), 
the final signal to noise level is roughly similar to the one of 
an XMM image of 95 Ks of
exposure time. The reference X-ray data for this simulation was
an image of the Lockman Hole field
 adquired in the 2-7 KeV energy band with the pn 
detector of the XMM-Newton satellite (with a pixel
size of  $4'' \times 4''$). The background map (mean value of $10^{-5}$
cts/s/pixel) was
estimated with the \emph{esplinemap} SAS v5.2 task. Simulation 3 was performed
so that the largest cluster size (in pixel units) and the signal to noise
were approximately equal to 
the conditions in the central region of the reference image.
The final signal to noise ratio is greater than in simulations 1 and
2. Figure~\ref{sim3} shows the simulated clusters (left panel), 
the complete simulation with noise (central panel) and the 
coefficient map that corresponds
to the simulation filtered with an scale-adaptive 
filter with $r_{c}=3.0$ pixels. 

The detection and determination of the amplitude and the scale of the 
clusters were
performed following the same steps than in simulation 2. The set of
chosen  `core radii' was $r_{c}=1.2,1.3,1.5,...,5.1$  pixels ($N=40$
filters). To consider a detection as a source it must be present in at least ten 
of the filtered maps (we chose this number 
for the same reasons that were explained in simulation 2).
The results are shown in table~\ref{tb5}. Simulation 3 has
two qualitative differences with respect to
simulations 1 and 2. 
First of all, the cluster profile drops much faster in
the X-ray case than in the SZ case. Second, 
the simulated background is white noise instead of $\propto
q^{-3}$. 
This is not the optimal situation since the gain achieved by the
scale-adaptive filter is not very high in comparison with other filters when
the background is dominated by white noise. In particular, the gains
obtained in simulation 3 are only $\sim 2$ (table~\ref{tb5}).
Moreover, 
the fast decline of the cluster profile makes them much more compact (that is,
more point-like), making more difficult to distinguish them from noise
fluctuations. 
In other words, comparatively to a microwave image of the same pixel scale and 
in which there 
are clusters of the same $r_{c}$, 
it is harder to estimate parameters such as
$r_{c}$ because the signal is condensed in a few pixels. 
If, for example, $r_{c}$ was
of the order or less than half a pixel, the cluster would be almost
indistinguishable 
from a point source.  This explains why, in spite of having the $r_{c}$ space
quite densely sampled, the mean error in the determination of that
parameter is almost 
of 0.5 pixels. There is no observed bias in the determination of
$r_{c}$. Clusters in  
simulation 2 had smaller core radii than those in simulation 3, and therefore
it was more 
likely to overestimate than to underestimate their size. 
In simulation 2,  smaller values of $r_{c}$ were chosen
in order to avoid overlapping among the clusters (the compactness of X-ray clusters reduces
the probability of overlapping).
Besides, the $P(q) \propto q^{-3}$
noise fluctuations grow stronger with their scale, so large-scale residuals (that, when combined
with the filtered profile, would lead to overestimation of the scale) were more likely in
simulation 2 that pixel-scale residuals (that, when combined
with the filtered profile, would lead to underestimation of the scale). Under white noise conditions
all scales of the background have the same power and therefore
there is no reason for bias in any of the two directions, as is observed.  

Figure~\ref{results3} shows the performance of scale-adaptive
filters in the determination
of the amplitude (bottom panel) and $r_{c}$ (top panel). 
The dispersion in the estimated $r_{c}$ 
is comparatively much greater than the dispersion in the estimated $A$
(lower panel of figure~\ref{results3}).
The dispersion in the estimation of the amplitude is
comparatively greater than in previous simulations due to the fact that
the dispersion of the filtered map (that is, the intrinsic 
statistical error in the determination of the amplitude)
is greater in simulation 3 than in simulation 1 and 2.
These errors, however, remain relatively small ($\sim 15\% $, table~\ref{tb5}).
The small relative errors in the determination of $A$ indicate 
that the estimated amplitude is quite stable
with respect to  
variations in $r_{c}$, that is, that curves $A$ versus
{$r_{c}$ are fairly flat indeed.  
The observed bias in the determination of the amplitude is around $5-10\%$,
similar to that observed in simulations 1 and 2.

Finally, the compactness of the clusters embedded in white noise has
the effect of 
increasing the number of spurious detections, especially for low
detection thresholds. At the $5\sigma$ level, however,
the number of spurious sources is less than $5\%$ of the true detections.
Therefore, the scale-adaptive filter is well suited to detect and extract
multiquadric profiles even in the less favorable case of more compact
clusters and a background dominated by white noise.

\section{Comparison with Wavelets} \label{sec_compa}

In order to compare the performance of the scale-adaptive 
filter with a wavelet-based analysis, we applied a Mexican Hat Wavelet
and the difference of two Gaussians (eq.~\ref{2gauss})
to one of the simulations described in the previous section.
Simulation 1
 (section~\ref{sec_sim1}). 
 was chosen for the sake of simplicity. 

The scale-adaptive filter, by construction, 
directly gives the amplitude of the sources after filtering. On
the contrary, the relation between the wavelet coefficients and the
amplitude of the sources has to be calculated (or, alternatively,
the wavelet has to be a priori re-normalized in order to  directly give
the amplitudes). Besides, the scale-adaptive filter 
for a multiquadric profile with a fixed scale is
a function of the
core radius $r_c$ 
(see eq. (\ref{SAFmq})) and therefore the variation of the scale
of the filter relates directly with $r_c$. The parameter $r_c$ in eq.  
(\ref{SAFmq}) corresponds exactly with the real $r_{c_o}$ of the clusters
so that, due to condition (iii), the coefficient $w(r_c,\vec{0})$ 
has a maximum when $r_c=r_{c_o}$.
The relation between the width of the Mexican Hat or the difference of
two Gaussians and $r_c$ is less clear and has to be analytically
or numerically calculated.
We suggest a procedure based on the calculation of 
the derivative of the wavelet coefficient with
respect to the wavelet width. 

  For this test we filtered the same simulation (simulation 1)
with three different
`filters': a scale-adaptive filter (see eq. (\ref{SAFmq})), a Mexican
Hat Wavelet (eq. (\ref{mexhateq})) and a wavelet generated
by the difference of two Gaussians (eq. (\ref{2gauss})). For the
scale-adaptive filter we repeated the same procedure described
in section~\ref{sec_sim2}, that is, we tried to determine
both the amplitude and the scale of the sources simultaneously. 
In the case of the wavelets, we performed an analogous 
analysis; instead of varying the parameter $r_c$ we varied the
widths $R$. 
The relation between $R$ at the maximum and $r_c$ was
analytically calculated for the Mexican Hat and the difference of
two Gaussians. In a realistic case (with a priori unknown power
spectrum of the noise) this relation must be calculated numerically.
The relation between the coefficient at the waveleth width 
$R$ corresponding with the
maximum and the amplitude $A$ was also analytically calculated.
These calculations are not necessary in the case of scale-adaptive
filter (they are included by construction).

The results of the test are showed in table~\ref{tb4_bis}.
   The results of this test were compared taking into account 
five different aspects: sensitivity (that is, number of detections),
reliability (proportion of spurious detections), gain, estimation
of the amplitude and estimation of the core radius.
Regarding the number of detections, there were no statistical
differences between the three filters. The number of spurious
detections, however, was significantly higher in the case of the difference
of two Gaussians (4 spurious detections versus 1 in the case of
the Mexican Hat and 0 in the case of the scale-adaptive filter). 
Comparing with table~\ref{tb3} we see that the number of spurious 
detections has dropped from 14 to 0 in the case of the scale-adaptive
filter. This is  due to 
the fact that we discard those `source candidates' that are not detected in
at least 5 filtered maps (as mentioned is section~\ref{sec_sim2}).
Regarding gain, scale-adaptive filter works better than the
two wavelets. In this simulation the number counts as
a function of the amplitude of the sources
is a constant, but in many realistic cases the number of
sources grow quickly as they become fainter; in that case, any
improvement in the gain of a filter will lead to a dramatic
increase of the number of detections.
Finally, the estimation of the amplitude and the radii of the
clusters is similar in the three  cases under study. The scale-adaptive
filter seems to give slightly smaller errors in the determination
of the core radii.

It is not difficult to explain the results of this test. The three
considered filters are able to remove the large-scale background
that dominates the image (the three of them drop to zero when
 $q\rightarrow 0$)
and therefore give good results in the detection. 
However, the scale-adaptive filter gives better reliability in
the detection and higher gains because its shape is better fitted
to the profile of the sources. The Mexican Hat
correlates well with Gaussian objects; if the
source profile departs from the Gaussian this correlation is not
perfect. Scale-adaptive filters are designed for a specific
source profile; if different kind of objects are present in an
image, a specific scale-adaptive filter can be constructed for
any of them. 
Regarding the estimation of parameters, in this example the 
three filters work similarly because they have been used under
the same philosophy. However, in the case of
the Mexican Hat and the difference of two Gaussians it was
necessary to find the relations between the coefficients and
the amplitudes, and between the width of the wavelet 
at the maximum detection and $r_c$. In the simplistic
case that has been considered these relations were straightforward
because they could be calculated analytically. In a realistic case
they should be calculated numerically, thus involving 
a maximization routine (with the consequent increase
in processing time and complexity). Scale-adaptive filter
does not suffer from this problem because the two mentioned 
relations are set equal to the identity 
by definition (conditions (i) and (iii)).

\section{Conclusions} \label{conclusions}

In this paper we have used the concept of scale-adaptive 
filter and applied it to a multiquadric profile
characterized by two parameters: the core radius $r_c$ and 
the the decay parameter $\lambda$, in
order to obtain an unbiased estimation of the amplitude of the source.
We have obtained explicit analytical formulas on Fourier space and 
simple analytical ones on real
space for some source profiles and backgrounds. In particular, 
we focus on the interesting cases of microwave and X-ray emissions. A
comparison with other standard filters is done. In particular, we
remark that for $\gamma = n$ 
(e. g. 1D signal and $1/f$ noise, 2D image and $P(q)\propto q^{-2}$)
the scale-adaptive and matched filters coincide.
Besides, a comparison with wavelet analysis techniques is performed.
The goals of scale-adaptive filters and wavelets are not the same: while
wavelets aim to fully describe structures in both space and
frequency domains, scale-adaptive filters focus only in the problem
of how to maximize the probability of detecting sources with a
given profile with high reliability. 
Scale-adaptive filters adopt some features of the  continuous 
wavelet transform, namely self-similarity and  the
ability to perform a sort of continuous multiresolution analysis. This
analysis at multiple scales can be used to estimate the scale of
sources and to improve the reliability of the detections.
In this context, wavelets are more flexible
and versatile, while filters are more powerful for a single yet
important task. 

We have simulated 2D clusters with $\lambda = 1/2$, with different
core radius $r_c$, embedded  
in a background $P(q)\propto q^{-3}$ that can mimic microwave emission
from the cluster plus 
intrinsic microwave and foreground emissions and noise. When the characteristic
scale (the `core radius' $r_c$) is known {\it a priori} we are able to recover 
a great number of sources without any significant error in their position and 
with errors in the determination of the amplitude of $\sim10\%$. About a $20\%$ of
spurious detections are also detected. 
This percentage of spurious sources could be reduced by introducing
a multiple-scale analysis 
 (i.e. imposing that
the sources appear in a certain number of maps filtered with different scales).
However, this would also reduce the number of true detections. 
The mean gain in this simulated case is
$4.4$, meaning that over a $5\sigma$ detection threshold in the filtered maps
we are able to detect sources that were only at $1.14\sigma$ in the original map.

When the characteristic scale is not known {\it a priori}, as will happen in
a realistic case, the size of the clusters can be estimated by filtering
the map with several filters at different scales and 
looking for the maximum among the coefficients  at the position of the sources.
Moreover, the multiscale analysis can be used to reject spurious detections
(that typically appear only in one or a few of the filtered maps) by
imposing that the source `candidates' must be present in several filtered maps. 
Unfortunately, the price of removing spurious detections is always to
reduce the number of detections. 
We performed a simulation including 
clusters of different sizes and applied a multiscale analysis
with ten filters of different scale. Only sources appearing in five or
more filtered maps were considered as true detections. Under these conditions, 
we were able to recover  $30\%$ less sources than in our first simulation, but
the number of spurious detections dropped to 0. The
position of the sources was again recovered with no significant error. 
The mean error in the determination of the amplitude is $< 15 \%$. Additionally, 
we were able to determine the scale parameter of the detected clusters with
mean errors of $\sim 0.15 $ pixels.   

We have also simulated 2D clusters with $\lambda = 3/2$, with different core radius $r_c$, 
embedded in a background $P(q) = constant$ 
that can mimic X-ray emission from the cluster plus
white noise. 
Once more, the multiscale analysis 
allows us to estimate the position, amplitude and scale of the sources
with small errors. The results are slightly worse than in the microwave simulations 
because the X-ray clusters are more compact and because the background
used in this simulation (white noise) is not the most favorable
for the scale-adaptive filters. Scale-adaptive filters take advantage of 
the fact that in most cases the power spectrum of the noise has a 
maximum at a scale that is different from the scale of the sources. 
In the white noise case the power spectrum is constant at all scales. In
other words, the scale-adaptive filter produced high gains in our microwave simulations
because the background showed strong large-scale features
that were removed efficiently by the scale-adaptive filters. In the
white noise case, the gain is only $g=2.0$. In spite of that, the errors in the
determination of the fundamental parameters remain small and the number 
of spurious detections over the $5\sigma$ detection threshold is less than 
$5\%$ of the number of true sources detected.

An additional point to address is `source
confusion', which is often a problem in deep images (X-ray or SZ maps)
with a relatively large beam (with the exception of Chandra). Crowded fields
are problematic for filtering as well as wavelet techniques. 
Source confusion can affect the performance of scale-adaptive
filters in two ways. In the first place,
the profile of two overlapped multiquadric profiles is, in general,  
not a multiquadric. Thus,
the correlation between the filter and the overlapped sources will not
be the ideal.
In the second place, 
scale-adaptive
filters work under the assumption that the  power spectrum $P(q)$ of
the data is not dominated by the sources we are trying to detect. In the 
case of one single source with no noise, for example, the scale-adaptive
filter tends to the null filter. In that case, however, there is no
point in filtering. Filtering is necessary in the low signal to noise
regime, where sources do not dominate the image.
Source confusion is not expected to be a problem in the future CMB Planck
mission.

We conclude that the scale-adaptive filter is well 
suited to detect and extract 
multiquadric profiles embedded in generic 
homogeneous and isotropic backgrounds.

Scale-adaptive filters can be constructed for any source profile $\tau(x)$.
If objects with different profiles are present, the image  
must be convolved with
a set of scale-adaptive filters adapted for each kind of object. 
If the objects have different and not \emph{a priori} known sizes
the image can be convolved with a family of scale-adaptive filters 
of the same functional form while varying their scale. The number of
scales probed with such family can be set by the user in order to
 sample the scale space with the desired precision. 
In this work spherical symmetry of the sources has been assumed, but
this is not a fundamental requirement 
of the method. Scale-adaptive filters can be straightforwardly
generalized in order to deal with non-symmetric 
profiles.

The filters presented in this work can be easily implemented and included
in more general separation/detection methods. For example, the frequency
dependence of the Sunyaev-Zel'dovich effect can be taken into account
to perform the detection of clusters from multichannel microwave maps. 
Such method will be presented in a future work. 

\acknowledgments

We thank Silvia Mateos,
Mar\'\i a Teresa Ceballos, Francisco Carrera 
and Patricio Vielva for useful suggestions and comments.
DH acknowledges support from a Spanish MEC FPU fellowship. 
RBB acknowledges financial support from the PPARC in the form of a
research grant.
RBB thanks the Instituto de F\'\i sica de
Cantabria for its hospitality during a stay in 2001.
We thank FEDER Project 1FD97-1769-c04-01, Spanish DGESIC Project 
PB98-0531-c02-01 and INTAS Project INTAS-OPEN-97-1192 for partial financial support.

\clearpage

\figcaption[fig1.ps]{Optimal scale-adaptive filters and matched filters (in Fourier space) 
associated to the $\beta=2/3$ profile for the cases
$\tau(x)=\frac{1}{(1+(x/rc)^{2})^{\lambda}}$, $\lambda=1/2$
(microwave) and $\lambda=3/2$ (X-ray) and
a homogeneous and isotropic background with power spectrum $P(q)
\propto p^{-\gamma}$.  The two panels in the left of the figure show
the optimal scale-adaptive filters for the cases
 $\gamma=0$ (solid line), $\gamma=1$ (dotted line), $\gamma=2$
(short-dashed line) and $\gamma=3$ (long-dashed line).
There is a degeneracy for $\gamma=1$ 
and $\gamma=2$ in the two considered cases. 
The two panels in the right of the figure show the matched filters for the
mentioned cases.
For the matched filter and the case case $\lambda=1/2$ and $\gamma=0$ 
the modified profile
$\tau(x)=N [ \frac{1}{\sqrt{r_{c}^{2}+x^{2}}}-\frac{1}{\sqrt{r_{v}^{2}+x^{2}}}]$,
$ N=\frac{r_{c}}{1-r_{c}/r_{v}}$
has been used  ($B=\frac{1}{4\pi N \ln(\frac{r_{c}+r_{v}}{2\sqrt{r_{c}r_{v}}})}$).
\label{filters}}

\figcaption[simulation1.ps]{Simulation 1. Left panel shows the
simulated `clusters'. Central 
panel shows the data to be analyzed (`clusters' plus $q^{-3}$ noise). Right panel shows
the coefficient map (after filtering). Only pixels above $3\sigma$
have been plotted in the right panel.
\label{sim1}}

\figcaption[results_sim1.ps]
{Results of the detection (over a $3\sigma$ threshold) in simulation 1. 
The top panel shows the estimated
versus the original amplitudes.
Points $A=A_{o}$ are given
by a dotted line.
For the sake of clarity, error bars in the estimation of the amplitudes have not
been represented; all of them have the same value that can be 
interpreted as the dispersion $\sigma$ of the filtered map. For this simulation,
$\sigma=0.064$. The theoretical value of $\sigma$ can be calculated using eq. (\ref{eq_sigma}).
The bottom panel shows the gain of the detected sources against
the original amplitude. The mean gain is given by the horizontal dotted line.
\label{results1}}

\figcaption[simulation2.ps]{Simulation 2. Left panel shows 
the simulated `clusters'. Central
panel shows the data to be analyzed (`clusters' plus $q^{-3}$ noise). 
Right panel shows
the coefficient map (after filtering with an scale-adaptive filter of $r_{c}=1.50$). 
\label{sim2}}

\figcaption[results_sim2.ps]
{Results of the detection (over a $3\sigma$ threshold) in simulation 2. 
The top panel shows the
estimated versus the original original core radii.
The dotted line corresponds to
$r_{c}=r_{c_{o}}$.
The bottom panel shows the estimated 
versus the original amplitudes.
For the sake of clarity, error bars in the 
estimation of the amplitudes have not
been represented; all of them have the same value that can be 
interpreted as the dispersion $\sigma$ of the filtered map. For this simulation,
$\sigma=0.071$.
The theoretical value of $\sigma$ can be calculated using eq. (\ref{eq_sigma}).
Points $A=A_{o}$ are given
by a dotted line.
\label{results2}}

\figcaption[simulation3.ps]{Simulation 3. Left panel shows 
the simulated `clusters'. Central
panel shows the data to be analyzed (`clusters' plus white noise). 
Right panel shows
the coefficient map (after filtering with an scale-adaptive filter of $r_{c}=3.0$). 
\label{sim3}}

\figcaption[results_sim3.ps]
{Results of the detection (over a $3\sigma$ threshold) in simulation 3. 
The top panel shows the
estimated versus the original original core radii.
The dotted line corresponds to
$r_{c}=r_{c_{o}}$.
The bottom panel shows the estimated 
versus the original amplitudes.
For the sake of clarity, error bars in the estimation of the amplitudes have not
been represented; all of them have the same value that can be 
interpreted as the dispersion $\sigma$ of the filtered map. For this simulation,
$\sigma=0.15$.
The theoretical value of $\sigma$ can be calculated using eq. (\ref{eq_sigma}).
Points $A=A_{o}$ are given
by a dotted line.
\label{results3}}

\clearpage

 \begin{table}
 \begin{center}
 \begin{tabular}{ c c c c }
 \tableline
 \tableline
$\lambda$ &
$\gamma$ &
$\tilde{\Psi}_{o}(q)$ &
$\Psi_{o}(x)$ \\
\tableline

$1/2$ & 0 & $\frac{2}{\pi}e^{-y}$  & $\frac{2}{\pi r_{c}^{2}} \frac{1}{(1+x^{2})^{3/2}}$ \\
$1/2$ & 1 & $\frac{2}{\pi}ye^{-y}$ & $\frac{2}{\pi r_{c}^{2}} \frac{2-x^{2}}{(1+x^{2})^{5/2}}$ \\
$1/2$ & 2 & $\frac{2}{\pi}ye^{-y}$ & $\frac{2}{\pi r_{c}^{2}} \frac{2-x^{2}}{(1+x^{2})^{5/2}}$ \\
$1/2$ & 3 & $\frac{2}{\pi} (3-y) y^{2} e^{-y}$  & $\frac{4}{\pi r_{c}^{2}} \frac{-12x^{4} + 21x^{2} -2}
{(1+x^{2})^{9/2}}$ \\
$3/2$ & 0 & $\frac{2}{\pi} (2y-1)  e^{-y}$  & $\frac{6}{\pi r_{c}^{2}} \frac{1-x^{2}}{(1+x^{2})^{5/2}}$ \\
$3/2$ & 1 & $\frac{4}{3\pi}  y^{2} e^{-y}$  & $\frac{4}{3\pi r_{c}^{2}} \frac{2-3x^{2}}{(1+x^{2})^{7/2}}$    \\
$3/2$ & 2 & $\frac{4}{3\pi}  y^{2} e^{-y}$  & $\frac{4}{3\pi r_{c}^{2}} \frac{2-3x^{2}}{(1+x^{2})^{7/2}}$ \\
$3/2$ & 3 & $\frac{4}{15\pi} (5-y) y^{3} e^{-y}$  & $ \frac{4}{3\pi r_{c}^{2}} 
\frac{9x^{6}-84x^{4}+88x^{2}-2}{(1+x^{2})^{11/2}}$   \\

 \tableline
 \end{tabular}
 \caption{\label{filtertab1} 
 Optimal scale-adaptive filters associated to the $\beta=2/3$ profile for the cases
$\tau(r)=\frac{1}{(1+(r/rc)^{2})^{\lambda}}$, $\lambda=1/2$ (microwave) and $\lambda=3/2$ (X-ray).} 

 \tablecomments{Col. (1): $\lambda$ . 
Col. (2): Background (noise) exponent $\gamma$.
Col (3): Filter in Fourier space ($y=qr_{c}$).
Col. (4): Filter in real space ($x=r/r_{c}$).}

 \end{center}
 \end{table}
\clearpage

 \begin{table}
 \begin{center}
 \begin{tabular}{ c c c c }
 \tableline
 \tableline
$\lambda$ &
$\gamma$ &
$\tilde{\Psi}_{o}(q)$ &
$\Psi_{o}(x)$ \\
\tableline

$1/2$ & 0 & $B\frac{e^{-qr_{c}}-e^{-qr_{v}}}{q} $ &  $\frac{B}{N} \tau(x)$ \\
$1/2$ & 1 & $\frac{1}{\pi}e^{-qr_{c}}$ & $\frac{1}{\pi r_{c}^{2}} \frac{1}{(1+x^{2})^{3/2}}$  \\
$1/2$ & 2 & $\frac{2}{\pi}(qr_{c})e^{-qr_{c}}$ &$ \frac{2}{\pi r_{c}^{2}} \frac{2-x^{2}}{(1+x^{2})^{5/2}}$   \\
$1/2$ & 3 & $\frac{2}{\pi}(qr_{c})^{2}e^{-qr_{c}}$ &$ \frac{2}{\pi r_{c}^{2}}\frac{2-3x^{2}}{(1+x^{2})^{7/2}}$   \\
$3/2$ & 0 & $\frac{2}{\pi}e^{-qr_{c}}$ & $\frac{2}{\pi r_{c}^{2}} \frac{1}{(1+x^{2})^{3/2}}$   \\
$3/2$ & 1 & $\frac{2}{\pi}(qr_{c})e^{-qr_{c}}$ & $\frac{2}{\pi r_{c}^{2}} \frac{2-x^{2}}{(1+x^{2})^{3/2}}$  \\
$3/2$ & 2 & $\frac{4}{3\pi}(qr_{c})^{2}e^{-qr_{c}}$ & $\frac{4}{3\pi r_{c}^{2}} \frac{2-3x^{2}}{(1+x^{2})^{3/2}}$  \\
$3/2$ & 3 & $\frac{2}{3\pi}(qr_{c})^{3}e^{-qr_{c}}$ &$ \frac{2}{\pi r_{c}^{2}} \frac{3x^{4}-24x^{2}+8}{(1+x^{2})^{3/2}}$  \\

 \tableline
 \end{tabular}
 \caption{\label{filtertab2} 
 Matched filters associated to the $\beta=2/3$ profile for the cases
$\tau(r)=\frac{1}{(1+(r/rc)^{2})^{\lambda}}$, $\lambda=1/2$ (microwave) and $\lambda=3/2$ (X-ray).
For the case $\lambda=1/2$ and $\gamma=0$ 
the modified profile
$\tau(r)=N [ \frac{1}{\sqrt{r_{c}^{2}+r^{2}}}-\frac{1}{\sqrt{r_{v}^{2}+r^{2}}}]$,
$ N=\frac{r_{c}}{1-r_{c}/r_{v}}$
has been used  ($B=\frac{1}{4\pi N \ln(\frac{r_{c}+r_{v}}{2\sqrt{r_{c}r_{v}}})}$).}

 \tablecomments{Col. (1): $\lambda$ . 
Col. (2): Background (noise) exponent $\gamma$.
Col (3): Filter in Fourier space ($y=qr_{c}$).
Col. (4): Filter in real space ($x=r/r_{c}$).}

 \end{center}
 \end{table}

\clearpage

 \begin{table}
 \begin{center}
 \begin{tabular}{ c c c c c c c  }
 \tableline
 \tableline
$\sigma$ &
detected &
spurious &
mean offset &
$\bar{b}_{A}$ &
$\bar{e}_{A}$ &
$\bar{g}$ \\
\tableline
3.0 & 80  & 14  & 0.0 & 7.7 & 10.7 & 4.4 \\
4.0 & 71  & 17  & 0.0 & 6.6 &  9.5 & 4.4 \\
5.0 & 59  & 12  & 0.0 & 7.2 &  9.7 & 4.4 \\
 \tableline
 \end{tabular}
 \caption{\label{tb3} 
 Detections in simulation 1. The number of simulated sources is 100. 
}
 \tablecomments{Col. (1): $\sigma$ detection level. 
Col. (2): Number of true 
detections above
the $\sigma$ threshold. Col (3): Number of spurious sources. Col. (4): Mean
offset in the position of the source (pixels). Col. (5): Mean relative bias
in the determination of the amplitude ($\%$), defined as the average of
$100 \times \frac{A-A_{o}}{A_{o}}$. 
Col. (6): Mean relative error in the determination of the amplitude ($\%$),
defined as the average of $100 \times \frac{\mid A-A_{o} \mid}{A_{o}}$ .
Col. (7): Mean gain (as defined in eq. (\ref{gaineq}))}

 \end{center}
 \end{table}

\clearpage

 \begin{table}
 \begin{center}
 \begin{tabular}{ c c c c c c c c c  }
 \tableline
 \tableline
$\sigma$ &
detected &
spurious &
mean offset &
$\bar{b}_{A}$ &
$\bar{e}_{A}$ &
$\bar{b}_{r_{c}}$ &
$\bar{e}_{r_{c}}$ &
$\bar{g}$ \\
\tableline
3.0 & 57  & 0  & 0.0 & 11.5 & 14.0 & -0.12 & 0.14 &  3.7 \\
4.0 & 49  & 0  & 0.0 &  7.3 & 10.4 & -0.13 & 0.14 &  3.5 \\
5.0 & 38  & 0  & 0.0 &  6.1 &  8.5 & -0.12 & 0.13 &  3.5 \\

 \tableline
 \end{tabular}

 \caption{\label{tb4} Detections in simulation 2. The number of simulated sources is 100.}

\tablecomments { Col. (1): $\sigma$ detection level. Col. (2): Number 
of true detections above the $\sigma$ threshold. 
Col (3): Number of spurious sources. Col. (4): Mean
offset in the position of the source (pixels). Col. (5): Mean relative bias
in the determination of the amplitude ($\%$), defined as the average of
$100 \times \frac{A_{i}-A_{o_{i}}}{A_{o_{i}}}$.
Col. (6): Mean relative error in the determination of the amplitude ($\%$),
defined as the average of $100 \times \frac{\mid A_{i}-A_{o_{i}} \mid}{A_{o_{i}}}$.
Col. (7): Mean bias in the determination of $r_{c}$ (in pixel units),
defined as the average of $\frac{r_{c_{i}}-r_{c_{o_{i}}}}{r_{c_{o_{i}}}}$.
 Col. (8): Mean 
absolute error
in the determination of $r_{c}$ (pin pixel units),
defined as the average of  $ \frac{\mid r_{c_{i}}-r_{c_{o_{i}}} \mid}{r_{c_{o_{i}}}}$.
Col. (9): Mean gain (as defined in eq. (\ref{gaineq}))}

\end{center}
 \end{table}

\clearpage

 \begin{table}
 \begin{center}
 \begin{tabular}{ c c c c c c c c c  }
 \tableline
 \tableline
$\sigma$ &
detected &
spurious &
mean offset &
$\bar{b}_{A}$ &
$\bar{e}_{A}$ &
$\bar{b}_{r_{c}}$ &
$\bar{e}_{r_{c}}$ &
$\bar{g}$ \\
\tableline
3.0 & 73  & 22  & 0.4 & 10.1 & 17.8 & 0.06 & 0.40 &  2.0 \\
4.0 & 63  & 10  & 0.3 & 6.2 & 14.9 & 0.03 & 0.39 &  2.0 \\
5.0 & 48  & 2  & 0.2 & 6.9 & 12.6 & 0.05 & 0.36 &  2.0 \\

 \tableline
 \end{tabular}

 \caption{\label{tb5} Detections in simulation 3. The number of simulated sources is 100.}

\tablecomments { Col. (1): $\sigma$ detection level. Col. (2): Number 
of true detections above the $\sigma$ threshold. 
Col (3): Number of spurious sources. Col. (4): Mean
offset in the position of the source (pixels). Col. (5): Mean relative bias
in the determination of the amplitude ($\%$), defined as the average of
$100 \times \frac{A_{i}-A_{o_{i}}}{A_{o_{i}}}$.
Col. (6): Mean relative error in the determination of the amplitude ($\%$),
defined as the average of $100 \times \frac{\mid A_{i}-A_{o_{i}} \mid}{A_{o_{i}}}$.
Col. (7): Mean bias in the determination of $r_{c}$ (in pixel units),
defined as the average of $\frac{r_{c_{i}}-r_{c_{o_{i}}}}{r_{c_{o_{i}}}}$.
 Col. (8): Mean 
absolute error
in the determination of $r_{c}$ (in pixel units),
defined as the average of  $ \frac{\mid r_{c_{i}}-r_{c_{o_{i}}} \mid}{r_{c_{o_{i}}}}$.
Col. (9): Mean gain (as defined in eq. (\ref{gaineq}))}

\end{center}
 \end{table}

\clearpage

 \begin{table}
 \begin{center}
 \begin{tabular}{ c c c c c c c c }
 \tableline
 \tableline
Filter &
detected &
spurious &
$\bar{g}$  &
$\bar{b}_{A}$ &
$\bar{e}_{A}$ &
$\bar{b}_{r_c}$ &
$\bar{e}_{r_c}$ \\

\tableline
SAF & 87 & 0 & 4.4 & -11.1 & 13.4 & -0.08 & 0.08 \\
MHW & 89 & 1 & 4.2 & -10.1 & 13.4 & -0.09 & 0.13 \\
D2G & 88 & 4 & 4.2 & -12.2 & 13.5 & -0.08 & 0.11 \\

 \tableline
 \end{tabular}
 \caption{\label{tb4_bis} 
 Comparison of the performances of the scale-adaptive filter
and two wavelets. The analysis was performed over 
detections at the $3\sigma$ detection level using
Simulation 1. 
}
 \tablecomments{Col. (1): Filter type (SAF=scale-adaptive filter,
MHW=Mexican Hat Wavelet, D2G=Difference of two Gaussians).
Col. (2): Number of true 
detections above
the $\sigma$ threshold. Col (3): Number of spurious sources. 
Col. (4): Mean gain (as defined in eq. (\ref{gaineq}))
Col. (5): Mean relative bias
in the determination of the amplitude ($\%$), defined as the average of
$100 \times \frac{A-A_{o}}{A_{o}}$. 
Col. (6): Mean relative error in the determination of the amplitude ($\%$),
defined as the average of $100 \times \frac{\mid A-A_{o} \mid}{A_{o}}$ .
Col. (7): Mean bias in the determination of $r_{c}$ (in pixel units),
defined as the average of $\frac{r_{c_{i}}-r_{c_{o_{i}}}}{r_{c_{o_{i}}}}$.
 Col. (8): Mean 
absolute error
in the determination of $r_{c}$ (in pixel units),
defined as the average of  $ \frac{\mid r_{c_{i}}-r_{c_{o_{i}}} 
\mid}{r_{c_{o_{i}}}}$.}

 \end{center}
 \end{table}


\begin{thebibliography}{}

\bibitem[]{} Cay\'on, L., Sanz, J. L., Barreiro, R. B., Mart\'\i nez-Gonz\'alez, E., Vielva, P., 
  Toffolatti, L., Silk, J., Diego, J. M. \& Arg$\ddot{u}$eso, F., 2000, MNRAS, 315, 757

\bibitem[]{} Damiani, F., Maggio, A., Micela, G. \& Sciortino, S., 1997, ApJ, 483, 350

\bibitem[]{}  Freeman, P.E., Kashyap, V.,  
 Rosner, R. \&  Lamb, D.Q., accepted for publication in Ap. J. Supp. (v. 138 Jan. 2002).

\bibitem[]{} Grebenev, S.A., Forman, W., Jones, C. \& Murray, S., 1995, ApJ, 445, 607.

\bibitem[]{} Haykin, S., 1996, `Adaptive filter theory', Prentice-Hall, New York.

\bibitem[]{} Herranz, D., Gallegos, J., Sanz, J.L. \&  Mart\'\i nez-Gonz\'alez, E., 2002,
MNRAS in press.

\bibitem[]{} Hobson, M.P., Jones, A.W., Lasenby, A.N. \& Bouchet, F.R., 1998, \mnras, 300, 1.

\bibitem[]{} Hobson, M.P., Barreiro, R.B., 
Toffolatti, L., Lasenby, A.N., Sanz, J.L., Jones, A.W., \& Bouchet, F.R. 1999,
\mnras, 306, 232. 

\bibitem[]{} Irwin, M.J., 1985, MNRAS, 214, 575.

\bibitem[]{} Kawasaki, W., Shimasaku, K., Doi, M. \& Okamura, S., 1998, A\&A, 130, 567
\bibitem[]{} Lazzati, D., Campana, S., Rosati, P., Panzera, M.R. \& 
Tagliaferri, G., 1999, ApJ, 524, 414.

\bibitem[]{} Odgen, R.T., `Essential wavelets for statistical applications
and data analysis', 1997, Birkh\"auser, Boston.

\bibitem[]{} Postman, M., Lubin, L. M., Gunn, J. E., Oke, J. B., Hoessel, J. G., Schneider, D. P.
  \& Christensen, J. A., 1996, ApJ, 615, 111
\bibitem[]{} Rosati, P., Della Ceca, R., Burg, R., Norman, C. \&
Giacconi, R., 1995, ApJ, 445, L11.   

\bibitem[]{} Sanz, J.L., Arg\"ueso, F., Cay\'on, L., Mart\'\i nez-Gonz\'alez, E.,
Barreiro, R.B. \& Toffolatti, L., 1999a, MNRAS, 309, 672.

\bibitem[]{} Sanz, J.L., Barreiro, R.B., Cay\'on, L., 
Mart\'\i nez-Gonz\'alez, E., Ruiz, G.A., D\'\i az, F.J., Arg\"ueso, F., 
Silk, J. \& Toffolatti, L., 1999b, A\&AS, 140, 99.

\bibitem[]{} Sanz, J. L., Herranz, D. \& Mart\'\i nez-Gonz\'alez, E., 2001, ApJ, 552, 484
 
\bibitem[]{} Slezak, E., de Lapparent, V. \& Bijaoui, A., 1993, ApJ, 409, 517
\bibitem[]{} Tegmark, M. \& 
Efstathiou, G. 1996, \mnras, 281, 1927.
\bibitem[]{} Tegmark, M. \&
 Oliveira-Costa, A. 1998, \apj, 500, 83. 

\bibitem[]{} Valtchanov, I., Pierre, M. \& Gastaud, R., 2001, A\&A, accepted

\bibitem[]{} Vielva, P., Mart\'\i nez-Gonz\'alez, E., Cay\'on, L., Diego, J. M., Sanz, J. L. \& 
  Toffolatti, L., 2001a, MNRAS, 326, 181.

\bibitem[]{} Vielva, P., Barreiro, R. B., Hobson, M., Mart\'\i nez-Gonz\'alez, E., Lasenby, A., 
  Sanz, J. L. \& Toffolatti, L., 2001b, MNRAS, 328, 1.

\bibitem[]{} Vikhlinin, A., Forman, W., Jones, C., \& Murray, S., 1995, 
ApJ, 451, 542.
\bibitem[]{} Vikhlinin, A., et al., 1998, ApJ, 451, 542.

\end{thebibliography}
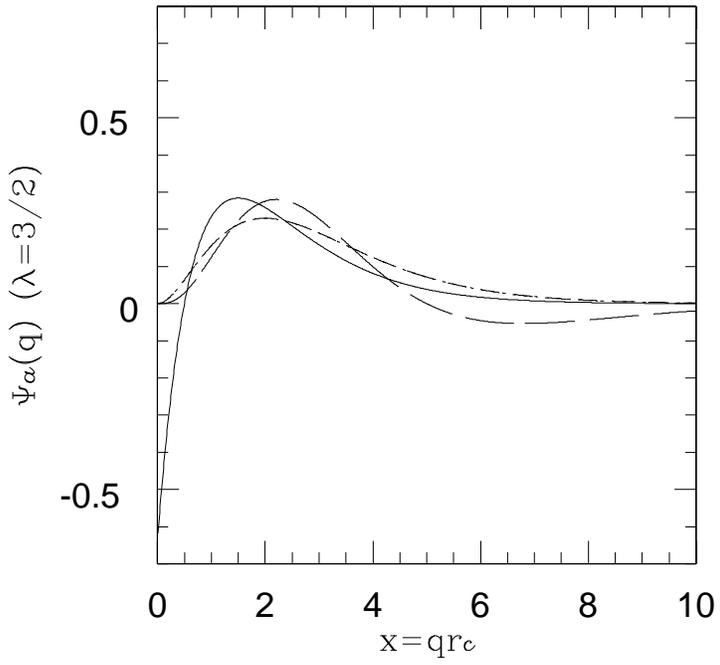
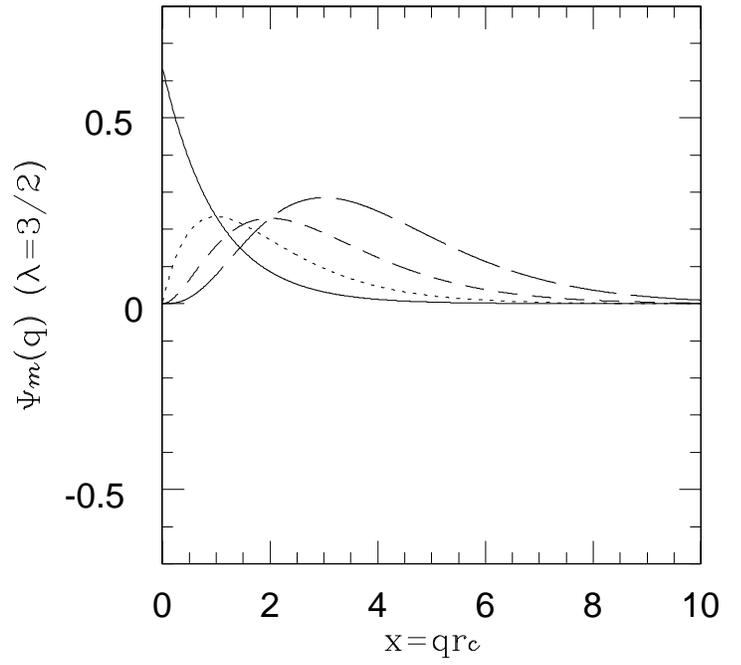
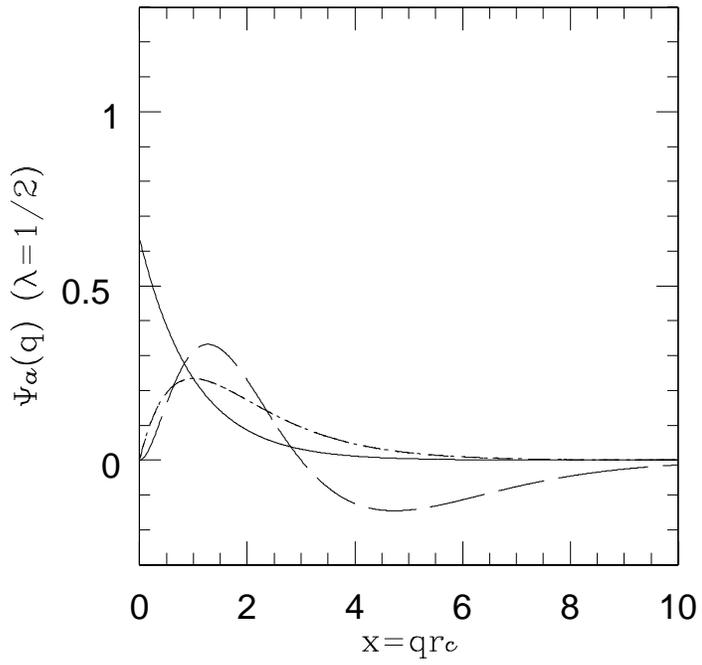
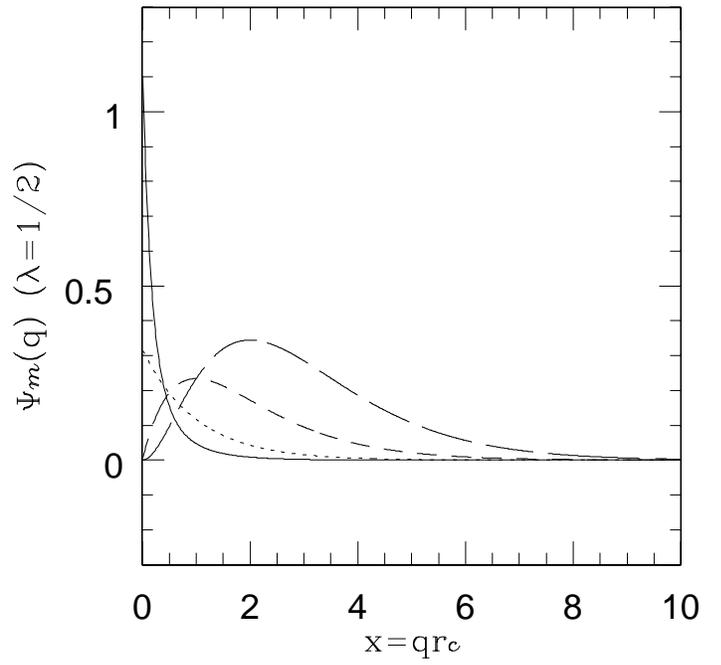
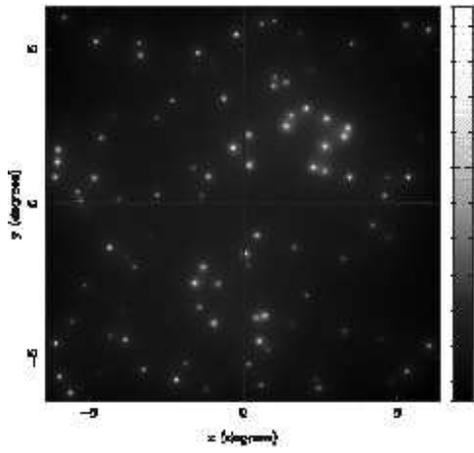
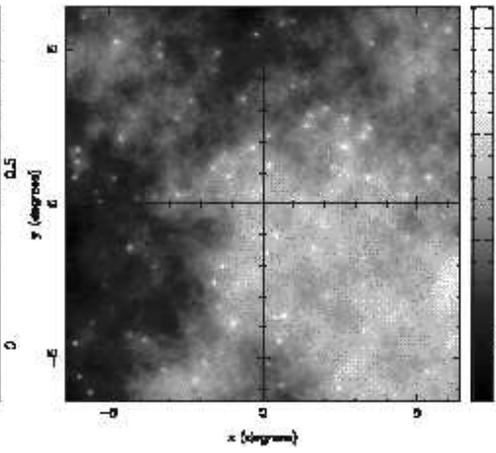
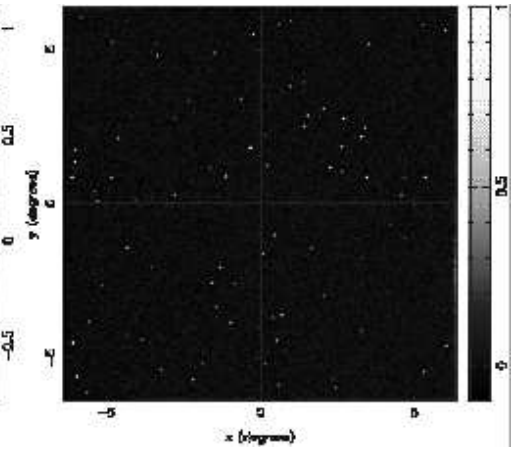
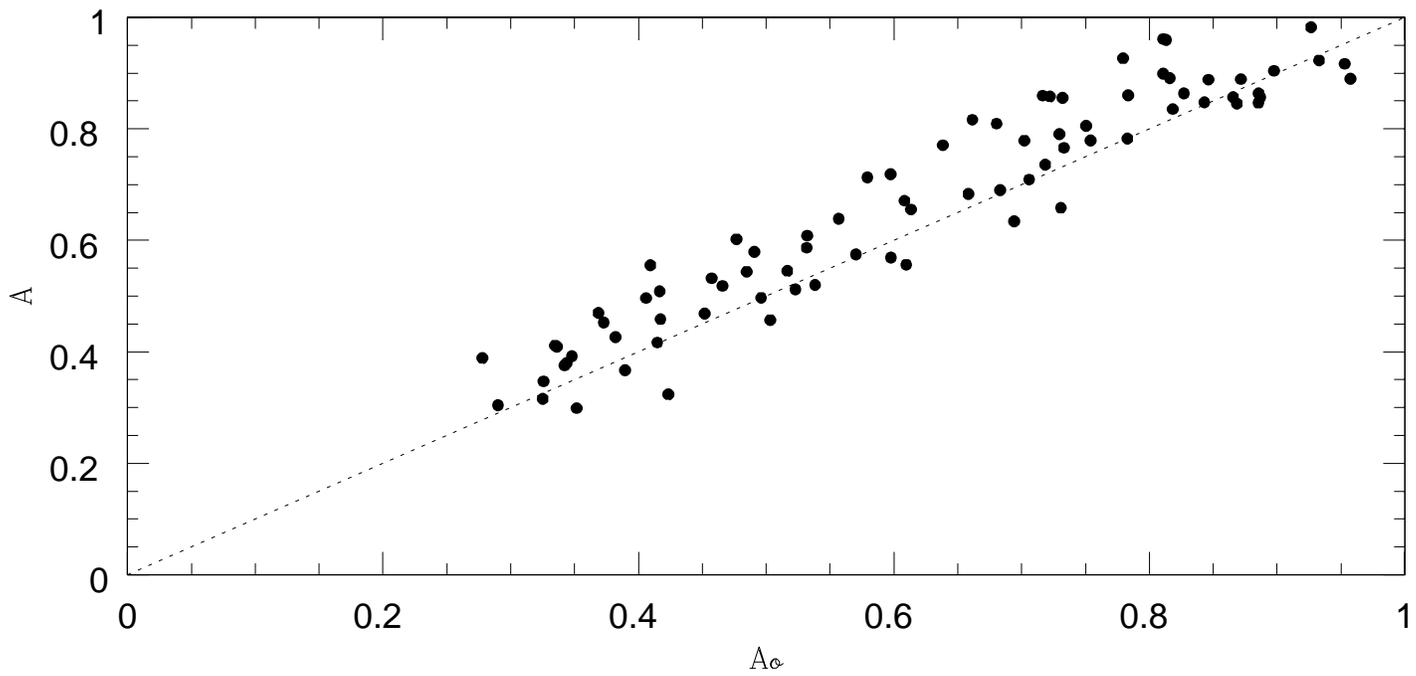
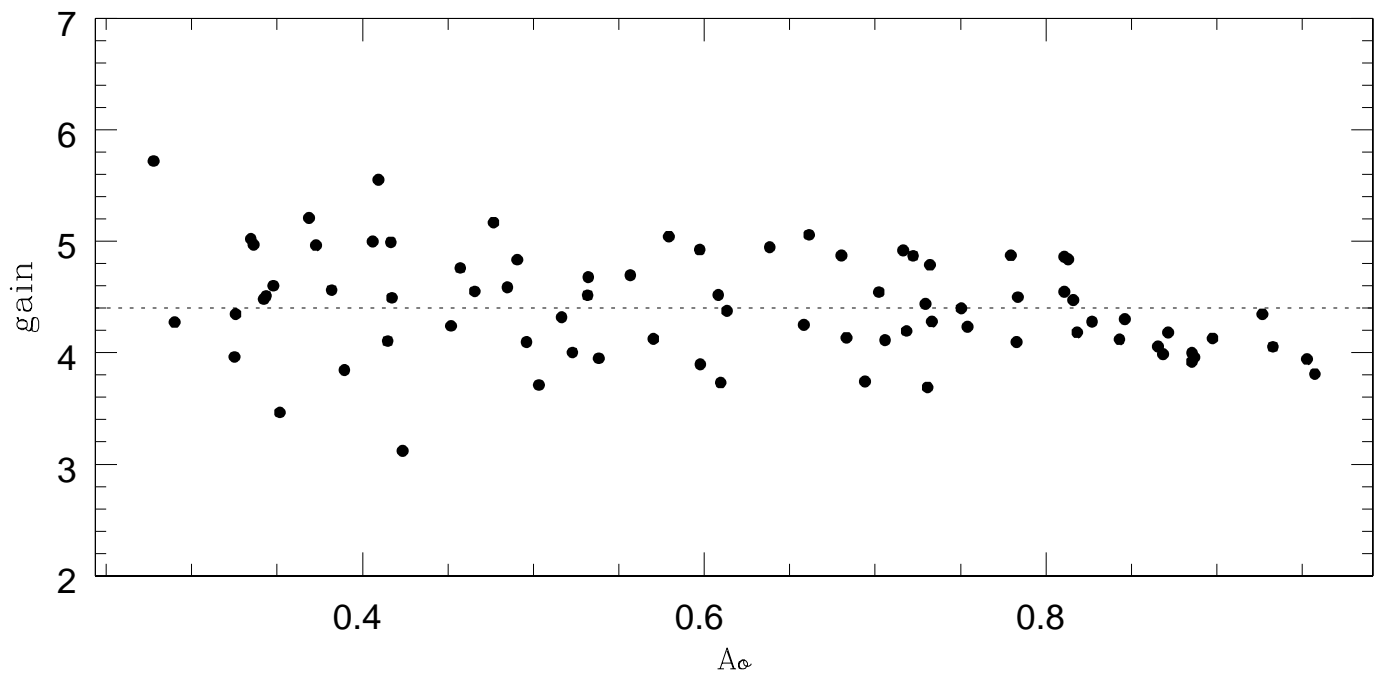
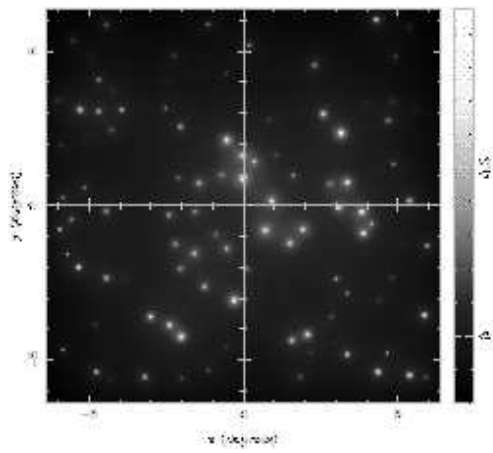
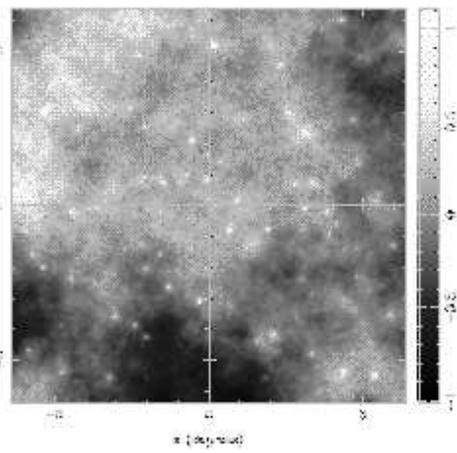
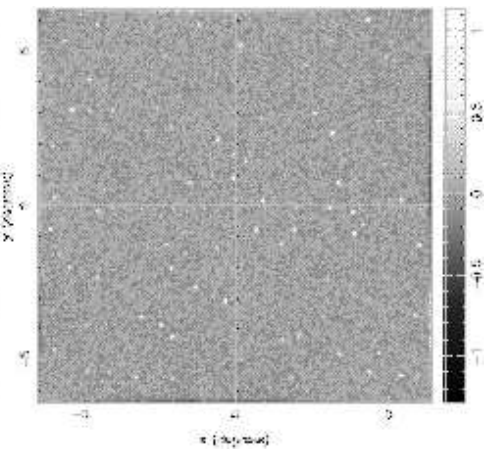
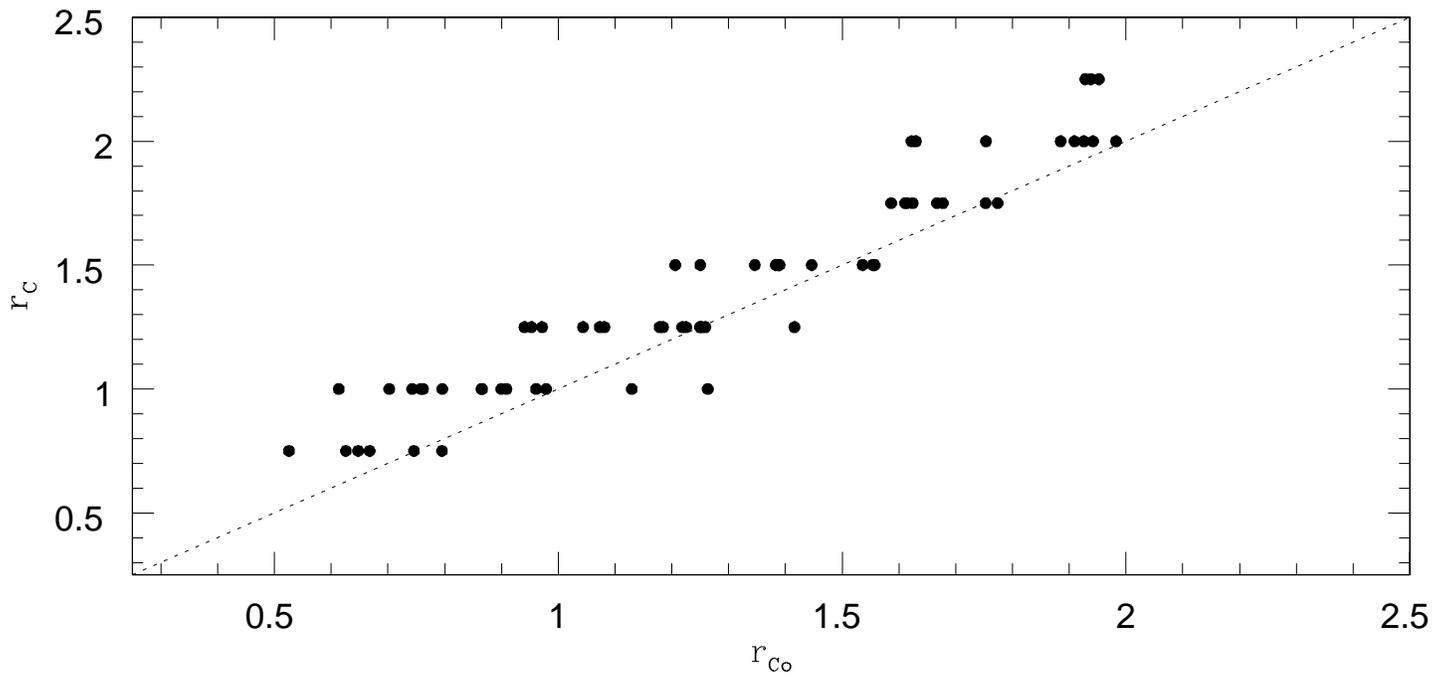
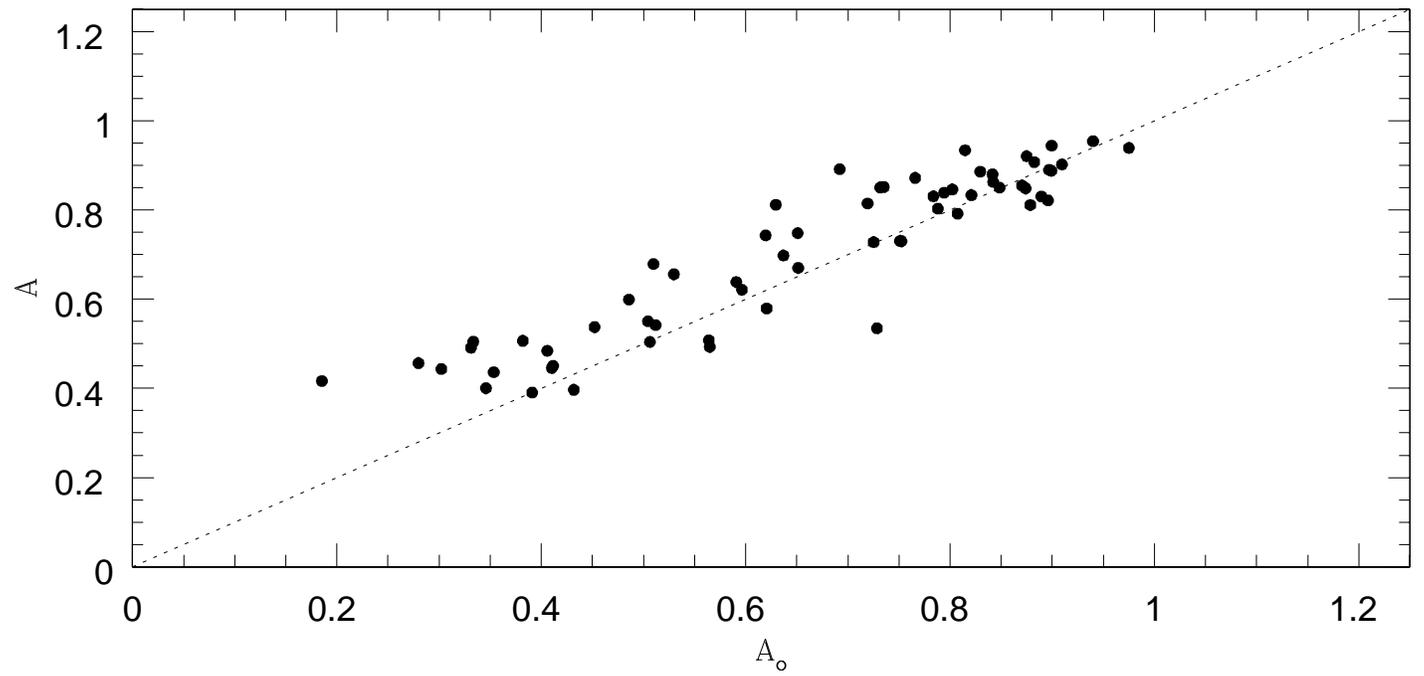
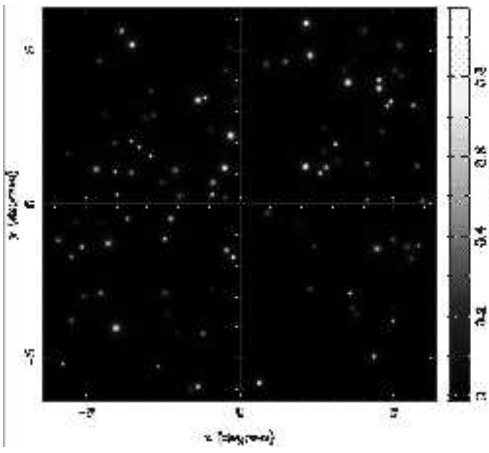
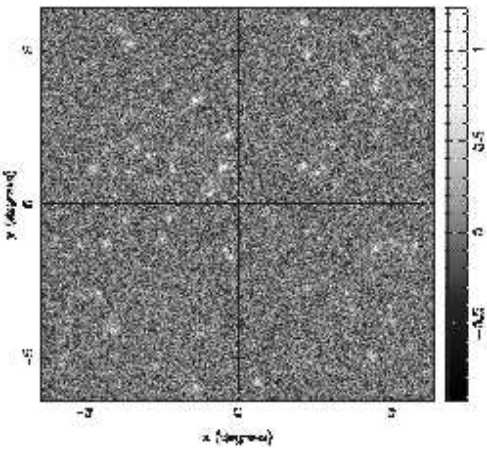
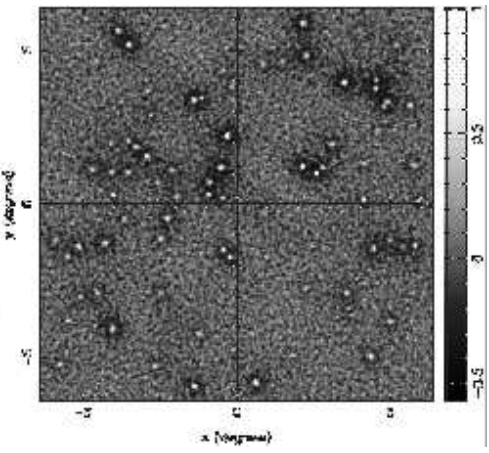
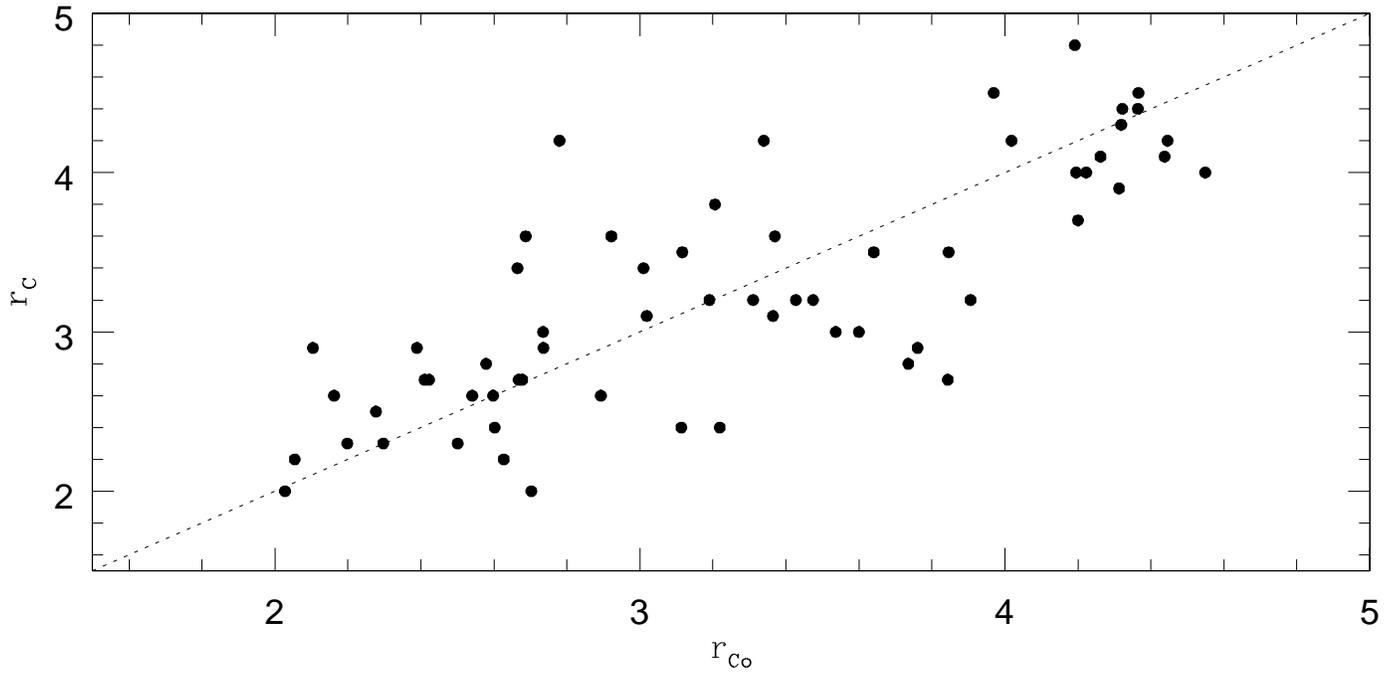
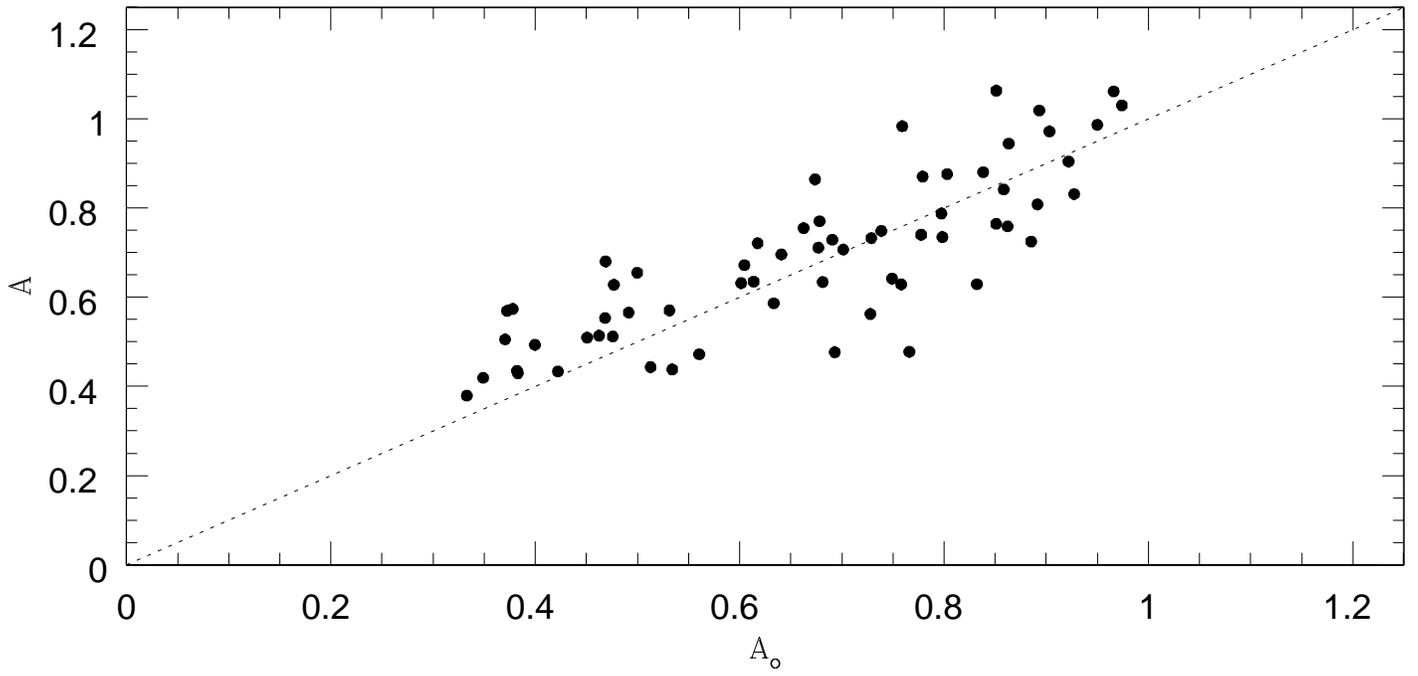
\end{document}